\documentclass[preprint,preprintnumbers,aps,nofootinbib,tightenlines,floatfix,showpacs]{revtex4}
\usepackage{amssymb,latexsym}
\usepackage{amsmath,amsbsy,bbm}
\usepackage{epsfig,bm}
\usepackage{graphicx,comment}
\usepackage{bm,comment}
\DeclareMathOperator{\tr}{tr}

\usepackage{epstopdf}
\usepackage{color}
\usepackage{subfigure}

\begin{document}
\def\a{{\alpha}}
\def\b{{\beta}}
\def\d{{\delta}}
\def\D{{\Delta}}
\def\X{{\Xi}}
\def\e{{\varepsilon}}
\def\g{{\gamma}}
\def\G{{\Gamma}}
\def\k{{\kappa}}
\def\l{{\lambda}}
\def\L{{\Lambda}}
\def\m{{\mu}}
\def\n{{\nu}}
\def\o{{\omega}}
\def\O{{\Omega}}
\def\S{{\Sigma}}
\def\s{{\sigma}}
\def\th{{\theta}}

\def\ol#1{{\overline{#1}}}

\def\Aslash{A\hskip-0.45em /}
\def\Dslash{D\hskip-0.65em /}
\def\Dtslash{\tilde{D} \hskip-0.65em /}

\def\CPT{{$\chi$PT}}
\def\QCPT{{Q$\chi$PT}}
\def\PQCPT{{PQ$\chi$PT}}
\def\tr{\text{tr}}
\def\str{\text{str}}
\def\diag{\text{diag}}
\def\order{{\mathcal O}}

\def\cF{{\mathcal F}}
\def\cS{{\mathcal S}}
\def\cC{{\mathcal C}}
\def\cB{{\mathcal B}}
\def\cT{{\mathcal T}}
\def\cQ{{\mathcal Q}}
\def\cL{{\mathcal L}}
\def\cO{{\mathcal O}}
\def\cA{{\mathcal A}}
\def\cQ{{\mathcal Q}}
\def\cR{{\mathcal R}}
\def\cH{{\mathcal H}}
\def\cW{{\mathcal W}}
\def\cM{{\mathcal M}}
\def\cD{{\mathcal D}}
\def\cN{{\mathcal N}}
\def\cP{{\mathcal P}}
\def\cK{{\mathcal K}}
\def\Qt{{\tilde{Q}}}
\def\Dt{{\tilde{D}}}
\def\St{{\tilde{\Sigma}}}
\def\cBt{{\tilde{\mathcal{B}}}}
\def\cDt{{\tilde{\mathcal{D}}}}
\def\cTt{{\tilde{\mathcal{T}}}}
\def\cMt{{\tilde{\mathcal{M}}}}
\def\At{{\tilde{A}}}
\def\cNt{{\tilde{\mathcal{N}}}}
\def\cOt{{\tilde{\mathcal{O}}}}
\def\cPt{{\tilde{\mathcal{P}}}}
\def\cI{{\mathcal{I}}}
\def\cJ{{\mathcal{J}}}

\def\eqref#1{{(\ref{#1})}}

\newcommand{\xpt}[0]{{$\chi$PT}}
\newcommand{\hbxpt}[0]{{HB$\chi$PT}}
\newcommand{\calE}{{\cal E}}
\newcommand{\calN}{{\cal N}}
\newcommand{\calO}{{\cal O}}
\newcommand{\calR}{{\cal R}}
\newcommand{\bfp}{{\bf p}}
\newcommand\half{{\textstyle{\frac{1}{2}}}}
\newcommand\fourth{{\textstyle{\frac{1}{4}}}}
\newcommand{\Hc}{\rm{h.c.}}
\newcommand{\Transpose}{T}
\newcommand{\sgn}{\rm{sgn}}
\newcommand{\Ns}{N_s}
\newcommand{\Nt}{N_\tau}
\newcommand{\UnitMatrix}{{\bf 1}}
\newcommand\beq{\begin{eqnarray}}
\newcommand\eeq{\end{eqnarray}}
\newcommand\Table[1]{Table~\ref{tab:#1}}
\newcommand\Fig[1]{Fig.~\ref{fig:#1}}
\newcommand\bal{\begin{align}}
\newcommand\eal{\end{align} }
\newcommand\eq[1]{Eq.~\ref{eq:#1}}
\newcommand\sect[1]{Sect.~\ref{sec:#1}}
\newcommand\refcite[1]{Ref.~\cite{#1}}
\newcommand\refscite[1]{Refs.~\cite{#1}}
\newcommand{\Det}{\mathop{\rm Det}}
\newcommand\lstate[1]{\langle{#1}|}
\newcommand\rstate[1]{|{#1}\rangle}
\def\epp{{\epsilon^{\prime}}}
\def\Dslash{D\!\!\!\!/}
\def\vslash{v\!\!\!/}

\preprint{LLNL-JRNL-522761}

\title{S-wave scattering of strangeness -3 baryons}

\author{Michael I. Buchoff}
\affiliation{%
Physical Sciences Directorate,
Lawrence Livermore National Laboratory,
Livermore, California 94550, USA}

\author{Thomas C. Luu}
\affiliation{%
Physical Sciences Directorate,
Lawrence Livermore National Laboratory,
Livermore, California 94550, USA}

\author{Joseph Wasem}
\affiliation{%
Physical Sciences Directorate,
Lawrence Livermore National Laboratory,
Livermore, California 94550, USA}



\begin{abstract}
We explore the interactions of two strangeness -3 baryons in multiple spin channels with lattice QCD.  This system provides an ideal laboratory for exploring the interactions of multi-baryon systems with minimal dependence on light quark masses.  Model calculations of the two-$\Omega^-$ system in two previous works have obtained conflicting results, which can be resolved by lattice QCD. The lattice calculations are performed using two different volumes with $L\sim2.5$ and $3.9$ fm on anisotropic clover lattices at $m_\pi \sim 390$ MeV with a lattice spacing of $a_s \sim 0.123$ fm in the spatial direction and $a_t\sim{a}_s/3.5$ in the temporal direction.  Using multiple interpolating operators from a non-displaced source, we present scattering information for two ground state $\Omega^-$ baryons in both the S=0 and S=2 channels.  For S=0, $k\cot\delta$ is extracted at two volumes, which lead to an extrapolated scattering length of $a^{\Omega\Omega}_{S=0}=0.16 \pm 0.22 \ \text{fm}$, indicating a weakly repulsive interaction.  Additionally, for S=2, two separate highly repulsive states are observed.  We also present results on the interactions of the excited strangeness $-3$, spin-$\frac{1}{2}$ states with the ground spin-$\frac{3}{2}$ states for the spin-1 and spin-2 channels. Results for these interactions are consistent with attractive behavior.
\end{abstract}
\maketitle

\section{Introduction}                                        

Lattice QCD calculations have advanced to the point that scattering phenomena for multi-hadron systems can be reliably calculated from first principles.  These calculations, performed through the analysis of two or more hadrons in a finite volume, allow for phase shifts and potential bound states to be studied non-pertubatively\cite{Luscher:1986pf,Luscher:1990ux}.   The majority of the focus of these calculations has been to explore baryon-baryon and meson-meson systems, where in the latter the scattering length for the I=2 $\pi\pi$ scattering system has been calculated to within a few percent\cite{Yamazaki:2004qb,Beane:2005rj,Beane:2007xs,Feng:2009ij,Dudek:2010ew,Beane:2011sc,Yagi:2011jn}.  Additionally, high precision calculations of bound states for the two lambda system\cite{Beane:2010hg,Inoue:2010es}, as well as the deuteron and $\Xi\Xi$ system\cite{Beane:2011iw} were performed recently.  In this work, we explore a different hyperon-hyperon system, namely the two strangeness -3 baryon ($\Omega\Omega$) system, where we present the scattering results for both ground and excited states.

While lattice QCD calculations of excited states using cubic irrep sources are well established, these lattice techniques have only been applied to mesonic scattering\cite{Dudek:2010ew} and have yet to be applied to two baryon scattering.  Single hadron excited states have seen a great deal of attention from the lattice community for both mesons\cite{Dudek:2010wm,Dudek:2011tt} and baryons\cite{Basak:2007kj,Bulava:2010yg,Edwards:2011jj}, where many states are consistent with their experimental counterparts.  Recent advances include calculations of the isoscalar meson spectrum\cite{Dudek:2011tt} with the use of the latest algorithmic methods for calculating disconnected diagrams\cite{Peardon:2009gh} and the employment of GPU technology\cite{Clark:2009wm}.  We extend this approach of utilizing multiple embeddings of lattice irreducible representations\cite{Basak:2005ir} to the two baryon system, with the ultimate goal being to extract higher partial wave interactions of nucleons from fundamental lattice calculations.  As a starting point, we apply these techniques to the $\Omega\Omega$ system in a relative s-wave state.

The lattice study of the $\Omega\Omega$ system is of interest for several reasons. Like most hyperon-hyperon systems, the interactions between two or more $\Omega^-$ baryons are poorly understood experimentally due to their large mass and relatively short lifetime.  To this end, lattice QCD calculations can predict phenomena in these systems and pinpoint signals for heavy ion scattering experiments, such as STAR or ALICE.  The $\Omega\Omega$ system has not received as much theoretical attention as its lighter hyperon counterparts, such as the H-dibaryon\cite{Jaffe:1976yi} and single $\Lambda$ hypernuclei\cite{Golak:1996hj,Nogga:2001ef,Kamimura:2008fx}.  However, within the last decade, this system was studied in the context of the chiral quark model\cite{Zhang:2000sv}, where it was found to prefer a bound ground state with a binding energy on the order of $100 \; \text{MeV}$.  A conflicting analysis\cite{Wang:1995bg} using the quark dislocation model finds the system to be weakly repulsive. Additionally, the interactions of the $\Omega\Omega$ system, along with the interactions between the $\Omega^-$ and other baryons, may prove to be relevant in dense systems several times nuclear density, such as the core of a neutron star\cite{Page:2006ud}.  These interactions may also play a role in the core of a supernova, ultimately determining whether the system becomes a neutron star or a black hole.

Another attractive aspect of studying the $\Omega\Omega$ system on the lattice is the fact that the system is believed to primarily depend on the physical strange quark mass as opposed to the unphysically large light quark masses with which these lattice calculations are performed.  This assertion is found to be true for the $\Omega^-$ and several of its excited states in Ref.~\cite{Bulava:2010yg} and, consequently, the $\Omega^-$ mass is often used to set the lattice scale\cite{Lin:2008pr}.  Thus, unlike most nuclear calculations involving light quarks, calculation of two-$\Omega^-$ systems at the physical point should rely less on chiral extrapolations.  Additionally, the inversions involving only the strange quark are less computationally expensive and the resulting signal involving strange baryons is cleaner.  For these reasons, the multi-$\Omega^-$ system is the ideal laboratory for understanding nuclear interactions on the lattice directly as it not only could provide insights into two or more nucleon interactions\cite{Beane:2003da,Beane:2009py,Inoue:2011ai,Beane:2011iw}, but it also provides a unique opportunity to study a host of nuclear interactions, such as the tensor forces and s-wave three-baryon forces\cite{Beane:2009gs,Yamazaki:2009ua}.

This work is organized as follows:  In \sect{two_omega}, the basic properties of the $\Omega\Omega$ system are mapped out along with differences from the two nucleon systems.  In \sect{two_omega_lattice}, the $\Omega\Omega$ system in a finite box  is explored, conventions are defined, and the methods of multiple embeddings are discussed.  In \sect{lattice}, the calculation details and analysis methods are explained, with the lattice results presented in \sect{single_omega_results} and \sect{two_omega_results}. Finally, in \sect{kcotdelta} the scattering results are derived, with a conclusion in \sect{concl}.

\section{Two $\Omega$ baryon system in infinite volume} \label{sec:two_omega}         
As in the case of two nucleons in the isospin limit, the channels with which two-$\Omega^-$ baryons can interact are restricted by the Pauli exclusion principle.  Each $\Omega^-$ contains three valence strange quarks and is spin-3/2 in its ground state. Pauli statistics dictate that the two $\Omega^-$ wavefunction must be antisymmetric.  Where the $\Omega\Omega$ system differs from two nucleons is each $\Omega^-$ is spin-3/2 as opposed to spin-1/2 and there is no isospin wavefunction.  In general, two spin-3/2 particles can exist in a total spin $S=$ 0, 1, 2, or 3 state.  However, the additional condition for an anti-symmetric wavefunction leads to the result that two $\Omega^-$ baryons in an s-wave state can only have two non-trivial spin channels: S=0 and S=2.

As mentioned previously, there have been two model calculations of the $\Omega\Omega$ system in the $S$=0 channel. The first model calculation in \refcite{Wang:1995bg} explores potential di-baryon systems via the quark dislocation and color screening model.   In this model, the quark model is generalized in several notable ways. Namely, the color screening/string tension interaction is included in the $Q$-$\overline{Q}$ potential along with a delocalized quark orbit in the wavefunction.   The authors find good agreement with the experimental N-N system.   For the $\Omega\Omega$ system, the authors find a weakly repulsive interaction\cite{Wang:1995bg}
\begin{equation}
\Delta E_{\Omega\Omega} = E_{\Omega\Omega} - 2M_\Omega = 43 \pm 18 \ \text{MeV}. \quad \quad \text{(Quark Disloc./Color-screen Model)}
\end{equation}
Here, the positive $\Delta E$ value implies the theory is not bound and likely weakly repulsive (near threshold).

The calculation in \refcite{Zhang:2000sv} explores the $\Omega\Omega$ system in the chiral $SU(3)$ quark model.  In essence, the chiral $SU(3)$ quark model generalizes the quark model, consisting of one-gluon exchange and a confining potential, to an $SU(3)$ sigma model in order to account for non-perturbative effects that affect the constituent quark mass.   The resulting Hamiltonian from the confining potential in this set up (whose 17 free parameters are determined from experiment) allows for two baryons systems to be studied by solving the resonating group method equations.  The authors find good agreement with experimental N-N and Y-N phase shifts, and find for the $\Omega\Omega$ system\cite{Zhang:2000sv}
\begin{equation}
\Delta E_{\Omega\Omega} = E_{\Omega\Omega} - 2M_\Omega = -116 \ \text{MeV}. \quad \quad \text{(}SU(3)\  \text{Chiral Quark Model)}
\end{equation}
The depth of this bound state is significant and could be detected in heavy ion experiments, as detailed in \refcite{Zhang:2000sv}.  Also, a binding energy of this magnitude would easily be resolved in state-of-the-art lattice QCD calculations by multiple standard deviations. Reference \cite{Zhang:2000sv} also points out potential issues with the quark delocalization model due to non uniform confinement potentials\cite{Yuan:1999pg}.  Ultimately, one would prefer a first-principle, model-independent lattice QCD calculation to address this debate.

\begin{figure}[!t]
\centering
\mbox{
\includegraphics[width=.25\columnwidth,angle=0]{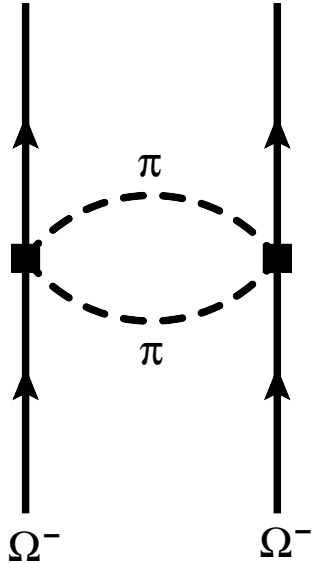}}
\caption{Leading diagram involving pion exchange for the $\Omega\Omega$ system.
\label{fig:pion interaction}}
\end{figure}

An over-arching goal of lattice studies of multi-baryon systems is to explore the connections between lattice calculations and parameters of many-body effective field theory.  This has been a primary goal for light baryons, but due to unphysically large quark masses this connection has proved difficult to achieve.  In the case of the multi-$\Omega^-$ system, the light quark mass dependence is expected to play a minimal role, as the leading order interaction involving pions is given by processes involving pair-produced two pion vertices\cite{Tiburzi:2008bk}, as depicted in \Fig{pion interaction}.   Consequently, lattice calculations with unphysical light-quark masses should provide `near-physical' results in the $\Omega$-only sector.  To date, there has not been much development in multi-$\Omega^-$ EFT due to the fact that low-energy physics of weakly decaying $\Omega^-$ baryons is difficult to probe physically.  However, with lattice QCD calculations, acquiring parameters for a meson-less EFT of multiple $\Omega^-$ baryons should be possible.  Recent work\cite{Ding:2012sm} has touched on this subject within the context of dark matter, and in a forthcoming paper we develop in detail the two flavor EFT for the two-$\Omega$ system.

In this work, we examine both ground state interactions and interactions of excited states in the s-wave two $\Omega^-$ system at a single pion mass of 390 MeV with two separate volumes.  Ultimately, we intend to probe tensor interactions by projecting the initial and final state baryons to higher partial waves. Future calculations will quantify to what extent the claim of minimal dependence on the light quark mass is valid.  Once understood, a new gateway to understanding interactions and systematics between EFT and lattice QCD calculations can be probed in a way currently inaccessible to light baryon systems.

\section{Two $\Omega$ system in a finite box} \label{sec:two_omega_lattice}         
There are several well-known complications when studying scattering calculations on the lattice.  First, discretized lattice calculations with periodic boundary conditions no longer preserve the $O(3)$ rotational symmetry observed in the continuum, but rather preserves an octahedral subgroup.  Thus, in order to extract information about the continuum S=0 and S=2 states, the appropriate irreducible representations of the octahedral group must first be understood.  Second, lattice calculations are performed in Euclidean space, where the usual LSZ formalism only holds at kinematic threshold\cite{Maiani:1990ca}.  As a result, phase shifts have to be extracted by analyzing energy shifts of hadrons in finite volume.  For two hadrons ($A$ and $B$) in a finite volume, the energies associated with the four-point correlation function are given by
\begin{equation}\label{eq:E_shift}
E = \sqrt{k^2+m_A^2} + \sqrt{k^2+m_B^2} = \Delta E + m_A + m_B,
\end{equation}
where $\Delta E$ is the energy of interaction and $k$ is the associated momenta in the center of mass frame.  As was shown in \refscite{Luscher:1986pf,Luscher:1990ux}, this interaction momenta at a given volume can be related to scattering phase shifts by
\begin{equation}\label{eq:Phase_Shift}
k\cot \delta(k) = \frac{1}{\pi L}S\left(\left(\frac{kL}{2\pi}\right)^2\right),
\end{equation}
where the function $S$ is given by the regularized three-dimensional zeta-function
\begin{equation}\label{eq:S}
S(\eta) = \sum_\mathbf{j \neq 0}^{|\mathbf{j}| < \Lambda} \frac{1}{|\mathbf{j}| ^2 - \eta}-4\pi\Lambda.
\end{equation}
The value of $k\cot \delta(k)$ extracted then has the normal effective range expansion given by
\beq\label{eq:effrange}
    k\cot \delta(k)&=&-\frac{1}{a}+\frac{1}{2}rk^2+...
\eeq
and given multiple values of $k\cot \delta(k)$ at multiple values of $k^2$ (obtained through lattice calculations of the same system at differing lattice volumes or through the use of boosted systems) one can make an extraction of the specific scattering parameters $a$, $r$, and so on. It is important to note that the relation in \eq{Phase_Shift} holds for elastic scattering processes and no longer holds when the energy of interaction exceeds twice the pion mass, while the relation in \eq{effrange} is only valid below the t-channel cut.  These facts often limit the number of excited scattering states that can be extracted from the lattice at a fixed volume.  The best way to understand this is through the realization that two non-interacting hadrons in a finite volume have quantized non-relativistic energy levels given by $4\pi n/(mL^2)$ for select integer values of $n$.  At a fixed volume, this reduces the access to excited scattering states for light hadrons, but allows for more excited scattering states to be explored for two $\Omega^-$ baryons.  Thus, for extracting excited behavior between multiple hadrons, including higher partial wave and tensor interactions,  the two $\Omega^-$ system is superior.  However, as will be discussed in more detail in the following sections, having more excited states accessible leads to more excited state contamination when trying to extract a given state.

In order to extract information about the $\Omega^-$, one must first calculate using an operator that has some (preferably large) overlap with the ground state.  Systems with definite integer spin modes in the continuum limit have dominant overlaps with different lattice irreducible representations (irreps),  labelled by $A_1$, $A_2$, $E$, $T_1$, and $T_2$.  The same can be said for fermionic modes which can map on to the lattice irreps $G_1$, $G_2$, or $H$.  Further, each representation $\Gamma$ will have an associated parity, which we label as $\Gamma^{\pm}$ for postive or negative parity. \Table{cubic irreps} enumerates the different lattice irreps used in this work and provides their dominant spin content in the infinite volume limit.
\begin{table}
\caption{Lattice irreps $\Gamma$ used in this work and dominant overlap with spin $J$ in the infinite volume limit ($L \rightarrow \infty$).  \label{tab:cubic irreps}}
\centering \begin{tabular}{c|c}
\hline
$\Gamma$ ($L\ne\infty$) & $J$ ($L=\infty$)   \\
\hline
\hline
$A_1^+$ & 0 \\
$T_1^-$ & 1 \\
$E^+$ & 2 \\
$T_2^+$ & 2 \\
\hline
$G_1^+$ & $\frac{1}{2}$\\
$H^+$ & $\frac{3}{2}$ \\
\hline
\hline
\end{tabular}
\end{table}
As seen from this table, the spin-$3/2$ $\Omega^-$ particle is represented by the $H^+$ irrep in finite cubic volumes.

The interpolating operator representing the $\Omega^-$ baryon is given by\cite{Basak:2005ir}
\begin{equation}\label{eq:Source_Def}
\ol \Omega_{\alpha \beta \gamma} = \epsilon_{a b c} \ol s^a_{\alpha} \ol s^b_{\beta}\ol s^c_{\gamma},
\end{equation}
where $a$, $b$, and $c$ are color indices and $\alpha$, $\beta$, and $\gamma$ are spinor indices.  Appropriate linear combinations of the spinor indices will produce $\Omega^-$ interpolating operators with definite lattice symmetry. For non-displaced sources and following \refcite{Basak:2005ir}, there are two representations, or embeddings, of the $\Omega^-$ particle in the $H^+$ irrep given by
\begin{eqnarray}\label{eq:ground_irrep}
{}^1H^+: \;
 \begin{array}{|c|c|}
 \hline
 \text{Interpolating operator}& | J, J_z\rangle\\
 \hline
 \hline
  \ol \Omega_{111} & |3/2,3/2\rangle\\
 \sqrt{3} \; \ol \Omega_{112} &  |3/2,1/2\rangle\\
 \sqrt{3}\; \ol \Omega_{122} & |3/2,-1/2\rangle\\
   \ol \Omega_{222} & |3/2,-3/2\rangle\\
 \hline
 \end{array}\quad
{}^2H^+: \;
 \begin{array}{|c|c|}
  \hline
 \text{Interpolating operator}& | J, J_z\rangle\\
 \hline
 \hline
 \sqrt{3}\;  \ol \Omega_{133} & |3/2,3/2\rangle\\
 2\ol \Omega_{134} + \ol \Omega_{233}&  |3/2,1/2\rangle \\
  \ol \Omega_{144}+ 2\ol \Omega_{234} &  |3/2,-1/2\rangle\\
  \sqrt{3} \;  \ol \Omega_{244} &  |3/2,-3/2\rangle\\
 \hline
 \end{array}. \nonumber
\end{eqnarray}
For a given embedding (${}^1H^+$ or ${}^2H^+$), each infinite volume $|J, J_z\rangle$-state and its corresponding source are given.  It is important to note that each state within a given embedding is orthogonal to the other states in that embedding after averaging over configurations.  Thus, each embedding can lead to four statistically independent calculations of the $\Omega^-$ two point function.  In the non-relativistic limit, the first embedding maps onto the upper two spinor components in the Dirac-Pauli basis, while the second maps onto the lower two components.  As such, one expects larger overlap with ground state systems when dealing with the first embedding.  It is also important to note that contracting the same state between two different embeddings is statistically dependent.

The s-wave states of the two $\Omega^-$ system can be formed from a tensor product of the ground state lattice irreps.  The lattice irrep that has overlap with the S=0 state is the $A_1^+$ irrep.  In terms of the spin-3/2 states of the $\Omega^-$ ground state, there is only one linear combination that leads to $A_1^+$ and it is given by\cite{Basak:2005ir}
\begin{equation}\label{eq:A1 }
A_1^+ (S=0): \quad \frac{1}{2}\Big(H_{\frac{3}{2}}  H_{-\frac{3}{2}} - H_{-\frac{3}{2}}  H_{\frac{3}{2}} +H_{-\frac{1}{2}}  H_{\frac{1}{2}}-H_{\frac{1}{2}}  H_{-\frac{1}{2}} \Big)
\end{equation}
where the subscript indicates the z-component of the spin and the $H$'s can be in either embedding.

For S=2, there are two lattice irreps that have overlap with the ground state, namely $E^+$ and $T_2^+$, and as a result both lattice irreps can be used as independent calculations of the S=2 $\Omega\Omega$ scattering state.  Additionally, unlike the $A_1^+$ lattice irrep, both $E^+$ and $T_2^+$ can be formed by multiple linear combinations of the two $H^+$ states, each of which are statistically independent determinations of the lattice irrep.  For $E^+$, there are two linear combinations given by
\begin{eqnarray}\label{eq:E }
E^+ (S=2)&:& \nonumber\\
1&:&\;\; \quad \frac{1}{2}\Big(H_{\frac{3}{2}}  H_{-\frac{3}{2}} - H_{-\frac{3}{2}}  H_{\frac{3}{2}} -H_{-\frac{1}{2}}  H_{\frac{1}{2}}+H_{\frac{1}{2}}  H_{-\frac{1}{2}} \Big) \nonumber\\
2&:&\;\; \quad \frac{1}{2}\Big(H_{\frac{3}{2}}  H_{\frac{1}{2}} - H_{\frac{1}{2}}  H_{\frac{3}{2}} -H_{-\frac{3}{2}}  H_{-\frac{1}{2}}+H_{-\frac{1}{2}}  H_{-\frac{3}{2}} \Big),
\end{eqnarray}
and for $T_2^+$, there are three linear combinations given by
\begin{eqnarray}\label{eq:T2 }
T_2^+ (S=2)&:& \nonumber\\
1&:&\;\; \quad \frac{1}{\sqrt{2}}\Big(H_{\frac{3}{2}}  H_{-\frac{1}{2}} - H_{-\frac{1}{2}}  H_{\frac{3}{2}} \Big) \nonumber\\
2&:&\;\; \quad \frac{1}{2}\Big(H_{\frac{3}{2}}  H_{\frac{1}{2}} - H_{\frac{1}{2}}  H_{\frac{3}{2}} -H_{-\frac{1}{2}}  H_{-\frac{3}{2}}+H_{-\frac{3}{2}}  H_{-\frac{1}{2}} \Big) \nonumber\\
3&:&\;\; \quad \frac{1}{\sqrt{2}}\Big(H_{\frac{1}{2}}  H_{-\frac{3}{2}} - H_{-\frac{3}{2}}  H_{\frac{1}{2}}  \Big).
\end{eqnarray}
It is worth noting that all linear combinations for $E^+$ and $T^+_2$ above will yield zero if each $H^+$ irrep is in the first embedding (the l.h.s. of Eq.~\eqref{eq:ground_irrep}).  Thus, to extract S=2 state, one minimally needs one $H^+$ irrep in the first embedding and the other in the second embedding, which naturally leads to a higher level of excited state contamination.

A good check worth pointing out is that the S=1 system should be trivial.  More specifically, the linear combinations that form the $T_1$ irrep should be zero due to parity restrictions and anti-symmetry.  Following Ref.~\cite{Basak:2005ir}, one such linear combination is given by $3(H_{3/2}  H_{-1/2} + H_{-1/2}  H_{3/2}) -4(H_{1/2}  H_{1/2})$.  Since each $H$ represents a source of three identical strange quarks, the exchange of any two $H$ terms will lead to an overall minus sign.  Thus, the first two terms will exactly cancel with each other and the last term can only be zero.  Thus, this linear combination will yield zero for the two $\Omega^-$ system.

In addition to the $H^+$ irrep that represents the $\Omega^-$ ground state, one can additionally explore excited states that correspond to spin-$1/2$ in the continuum limit.  The lattice irrep associated with this excited mode is $G_1^+$, which is given in terms of the $\Omega^-$ operators
\begin{eqnarray}\label{eq:excited_irrep}
G_1^+: \;
 \begin{array}{|c|c|}
   \hline
 \text{Interpolating operator}& | J, J_z\rangle\\
 \hline
 \hline
  \ol \Omega_{134}-\ol \Omega_{233}& |1/2,1/2\rangle\\
\ol \Omega_{144}-\ol \Omega_{234} &  |1/2,-1/2\rangle\\
 \hline
 \end{array}. \nonumber
\end{eqnarray}
Since the $H^+$ baryon and the $G_1^+$ baryon are not identical particles, scattering between these states can take all the spin values allowed by the addition of angular momenta (S=1,2).   The s-wave S=1 contribution (in irrep $T_1^+$) is given by
\begin{eqnarray}\label{eq:T1 }
T_1^+ (S=1)&:& \nonumber\\
1&:&\;\; \quad \frac{1}{2}\Big(G_{\frac{1}{2}}  H_{\frac{1}{2}} -3 G_{-\frac{1}{2}}  H_{\frac{3}{2}} \Big) \nonumber\\
2&:&\;\; \quad \frac{1}{\sqrt{2}}\Big(G_{\frac{1}{2}}  H_{-\frac{1}{2}} - G_{-\frac{1}{2}}  H_{\frac{1}{2}}  \Big) \nonumber\\
3&:&\;\; \quad \frac{1}{2}\Big(3G_{\frac{1}{2}}  H_{-\frac{3}{2}} - G_{-\frac{1}{2}}  H_{-\frac{1}{2}}  \Big)
\end{eqnarray}
and the s-wave S=2 contribution (in irrep $E^+$) is given by
\begin{eqnarray}\label{eq:E+ }
E^+ (S=2)&:& \nonumber\\
1&:&\;\; \quad \frac{1}{\sqrt{2}}\Big(G_{\frac{1}{2}}  H_{-\frac{1}{2}} + G_{-\frac{1}{2}}  H_{\frac{1}{2}} \Big) \nonumber\\
2&:&\;\; \quad \frac{1}{\sqrt{2}}\Big(G_{\frac{3}{2}}  H_{-\frac{1}{2}} + G_{-\frac{1}{2}}  H_{-\frac{3}{2}}  \Big).
\end{eqnarray}

Finally, it is possible to couple two $G_1$ excited $\Omega$ particles to form states of definite $A_1$ symmetry, corresponding to an s-wave, S=0 system in the infinite volume limit, using the coupling coefficients given in \refcite{Basak:2005ir}.   However, as we point out below, because of the high-energy levels associated with this system, coupled with limited statistics, we were not able to extract any statistically meaningful information from this system on the current lattices that were used in this work.

\section{Lattice Details}  \label{sec:lattice}          
\subsection{Configurations}
Our calculations were performed on anisotropic Wilson lattices generated on the uBGL machine at LLNL using the tuning parameters defined in \refcite{Lin:2008pr}. The primary ensemble used was a $20^3 \times 256$ with $m_\pi \approx 390$ MeV, $a_s \sim .1227$ fm, and $a_s/a_t \sim 3.5$ (see \refcite{Beane:2011iw} for more details of the anisotropic parameters).  The spatial extent of these lattices is $L \sim 2.5$ fm. On the $20^3 \times 256$ ensemble, 50 propagators with random sources were calculated on every 5 trajectories, where every measurement required one propagator. The propagators were generated on the Edge GPU cluster at LLNL. For the S=0 two-$\Omega$ system, we also performed measurements on $m_\pi \approx 390$ MeV lattices at a larger volume ($32^3 \times 256$) in an attempt to quantify volume effects.  Here 25 measurements per configuration, blocked every 4 trajectories, were made.  Table~\ref{table:configs} gives details of the configurations and measurements used for this work.

\begin{table}
\caption{Gauge configuration details\label{table:configs}}
\centering \begin{tabular}{|c|c|c|c|c|c|}
\hline
Size & $m_l$ &$m_s$ & $m_\pi$ [MeV] & $m_\pi L$ & Configs $\times$ Meas/Config   \\
\hline
$20^3 \times 256$ & $-0.840$ & $-0.743$ & $\sim 390$ MeV & 4.855&$1155 \times 50$ \\
\hline
$32^3 \times 256$ & $-0.840$ & $-0.743$ & $\sim 390$ MeV & 7.74&$465 \times 25$ \\
\hline
\end{tabular}
\end{table}

\subsection{Contractions}
\begin{figure}[!ht]
\centering
\mbox{
\subfigure[]{\includegraphics[width=.45\columnwidth,angle=0]{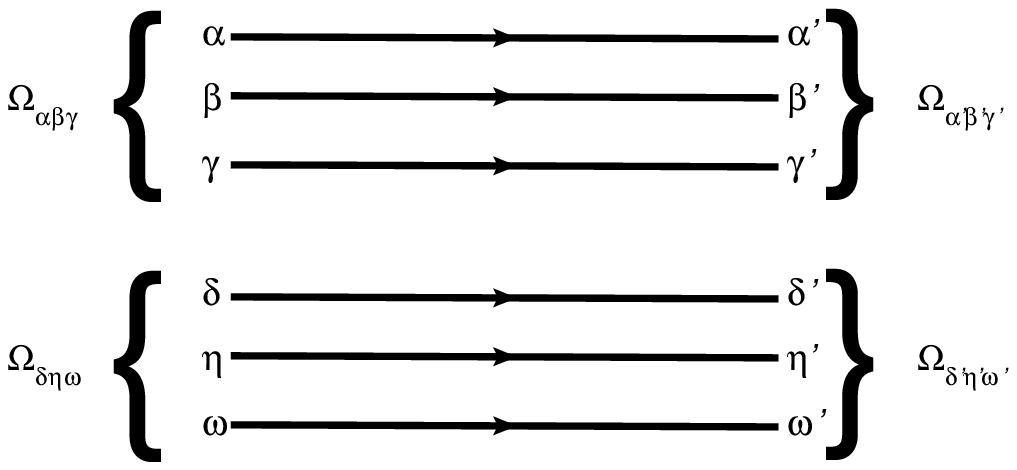}}\quad\subfigure[]{\includegraphics[width=.45\columnwidth,angle=0]{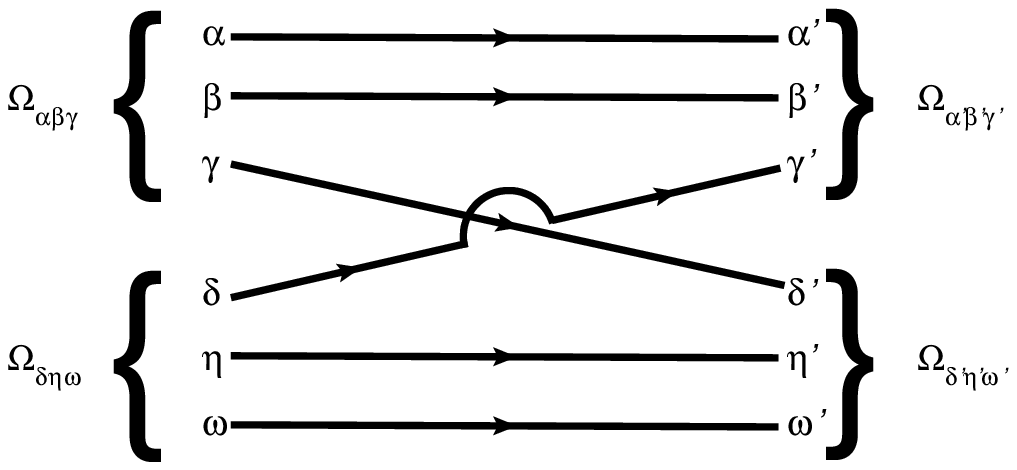}}}
\caption{Quark contractions types used in the calculation including the (a) direct contributions and the (b) exchange contributions.
\label{fig:contractions}}
\end{figure}

Due to the three degenerate valance quarks, the $\Omega^-$ interpolating operator has several symmetries worth noting.  The first symmetry is that the spinor indices can be interchanged freely,
\begin{equation}\label{eq:Sym1}
\ol \Omega_{\alpha \beta \gamma} = \ol \Omega_{ \beta \alpha \gamma} = \ol \Omega_{ \beta  \gamma \alpha}=\cdots .
\end{equation}
This fact is the result of exchanging two quarks leads to both a minus sign from permuting Grassman number and a minus sign from exchanging indices in the epsilon tensor, which cancel.  As a result, all Wick contractions will have the same relative sign.  For the two-$\Omega$ system, it can be shown using these symmetries that all contractions fall into two distinct forms: `direct'  and `exchange', as shown in \Fig{contractions}.
Taking advantage of these symmetries drastically reduces the 6!=720 possible contractions of two-$\Omega$ system by an order of magnitude and therefore reduces both computational cost due to matrix multiplication and required memory to hold the matrices.

Before performing the contractions, we first calculate the propagator using a Gaussian smeared source.  Upon inverting on that source, we Gaussian smear the sink (referred to hereafter as shell-shell, or SS measurements) or perform no smearing to the sink (referred to hereafter as shell-point, or SP measurements).  While the SS measurements are expected to give the best overlap with the ground state spectrum, by utilizing the SS and SP measurements in combination, one can largely eliminate the contribution from the first excited state and extract a more robust ground state signal that dominates at earlier Euclidean time\cite{Beane:2009kya,Luscher:1990ck}.  One can further enhance the ground state signal by making use of different combinations of the embeddings\footnote{Previous studies\cite{Basak:2007kj,Lin:2008pr,Bulava:2009jb,Bulava:2010yg,Edwards:2011jj} have used a matrix of correlation functions from different embeddings to extract the low-lying spectrum of several baryons, up to the first few excited states. As we are primarily concerned with the ground state baryons and their interactions these techniques are not used here.} discussed in \sect{two_omega_lattice}.  Using the propagator, we form the relevant irrep ``blocks"  for a given embedding, and we further suppress the excited states by projecting the momentum of both individual irrep blocks to zero independently, which will result in a correlation function where excited states with nonzero back-to-back momentum have been removed\cite{Beane:2009py}.  Finally, these irrep blocks are contracted and the correlation functions of interest are obtained.

\subsection{Analysis Details}
The measurements are blocked by configuration and the resulting ensemble is bootstrapped, with each bootstrap measurement and the final bootstrap ensemble being the same size as the original ensemble. For each embedding correlation function, the SS and SP measurements are put through a matrix-Prony algorithm as detailed in \refcite{Beane:2009kya}. Specifically (following the notation and derivation of \refcite{Beane:2009kya}), the correlation function recursion relation
\beq
    My_{\Gamma}(t+t_J)-Vy_{\Gamma}(t)&=&0
\eeq
has as a solution
\beq\label{eq:MPmatrix}
    M&=&\left[\sum_{t=\tau}^{\tau+t_W}y_{\Gamma}(t+t_J)y_{\Gamma}(t)^T\right]^{-1}, \ \ \ \ \ \ V=\left[\sum_{t=\tau}^{\tau+t_W}y_{\Gamma}(t)y_{\Gamma}(t)^T\right]^{-1}
\eeq
for the vector of correlation functions $y_{\Gamma}(t)$, of irrep and embedding type $\Gamma$.  Here, the window of timeslices from $\tau$ to $\tau+t_W$ is the set of values of the correlation function over which the outer product is taken. This window must include enough information to make the resulting matrix full-rank and invertible, with subsequent timeslices helping to reduce statistical noise to some extent\cite{Beane:2009kya}. The choice of $t_J$ will increase the ``lever-arm" that the matrix-Prony rotation provides, further mitigating statistical noise but at the price of increasing systematic fluctuations in the correlation function. As such this integer quantity should typically be chosen to be small. The eigenvectors $q_{\Gamma}$ of the matrix $V^{-1}M$, are defined by
\beq
    V^{-1}Mq_{\Gamma,i}=\lambda_{\Gamma,i}q_{\Gamma,i}
\eeq
where the eigenvalues $\lambda_{\Gamma,i}$ are placed in ascending order. This will return a correlation function $q_{\Gamma,0}$ that has an enhanced ground state contribution. Using the rotation matrix defined by $q_{\Gamma}$ on each of the bootstrap measurements for type $\Gamma$ will lead to an effective mass plot of the function
\beq
    M_{{\rm eff},\Gamma}(t)&=&\frac{1}{t_J}{\rm log}\left(\frac{q_{\Gamma,0}(t)}{q_{\Gamma,0}(t+t_J)}\right)
\eeq
with a longer and more robust plateau region. For each of the effective mass plots below, the values $t_W=10$ and $t_J=2$ have been used, with different rotation points $\tau$ for each type $\Gamma$. A fully correlated $\chi^2$ minimizing fit is then performed in the plateau region on the ensemble of bootstrapped effective mass data to extract the ground state energy with a statistical error. A systematic error from the fit window choice is obtained by modifying the endpoints of the fit window $\pm2$ timeslices and taking one-half the maximum minus the minimum of those fit values. The fit values are displayed along with the $\chi^2/dof$ for the fit and the $Q$ (or quality of fit) value, which is the integrated probability distribution of $\chi^2$ from the observed fit $\chi^2$ to infinity.

With this (and other black-box) methods there is serious concern that a false plateau may be recovered, as competing overlap factors under the rotation may produce a temporary cancelation in the plateau region that mimics the behavior of a real plateau. This behavior may be particularly problematic for methods with a large number of correlation functions in the vector $y_{\Gamma}$, as the opportunities for cancelations of the wrong form increase. To remove this complication, in this work all plateaus are initially identified in the effective mass fit of the SS data for each irrep and embedding $\Gamma$. This correlation function is manifestly positive (up to an overall sign) and so does not suffer from possible overlap cancelations. The matrix-Prony rotation point $\tau$ is then chosen such that $q_{\Gamma,0}$ returns an effective mass plateau fit value within 0.5$\sigma$ of the SS plateau fit value. The matrix-Prony result will then return a value that is statistically the same as the real plateau, but with a significantly improved signal-to-noise ratio.

\section{Strangeness -3 systems}\label{sec:single_omega_results}

\subsection{$s=\frac{3}{2}$: $H^{+}$ Irrep} \label{sec:H+}         
\begin{figure}[!ht]
\centering
\mbox{
\subfigure[]{\includegraphics[width=.5\columnwidth,angle=0]{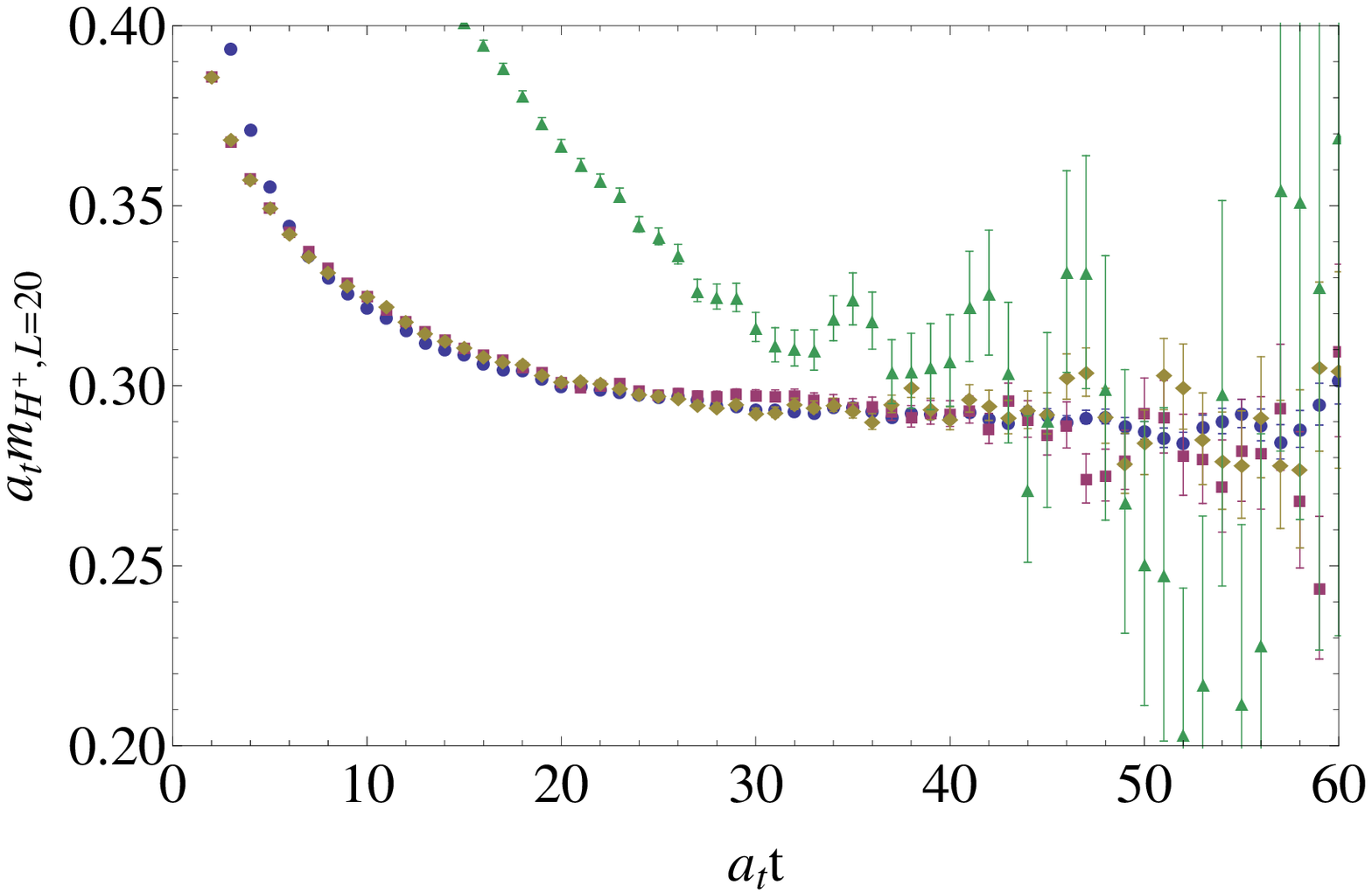}}\quad\subfigure[]{\includegraphics[width=.5\columnwidth,angle=0]{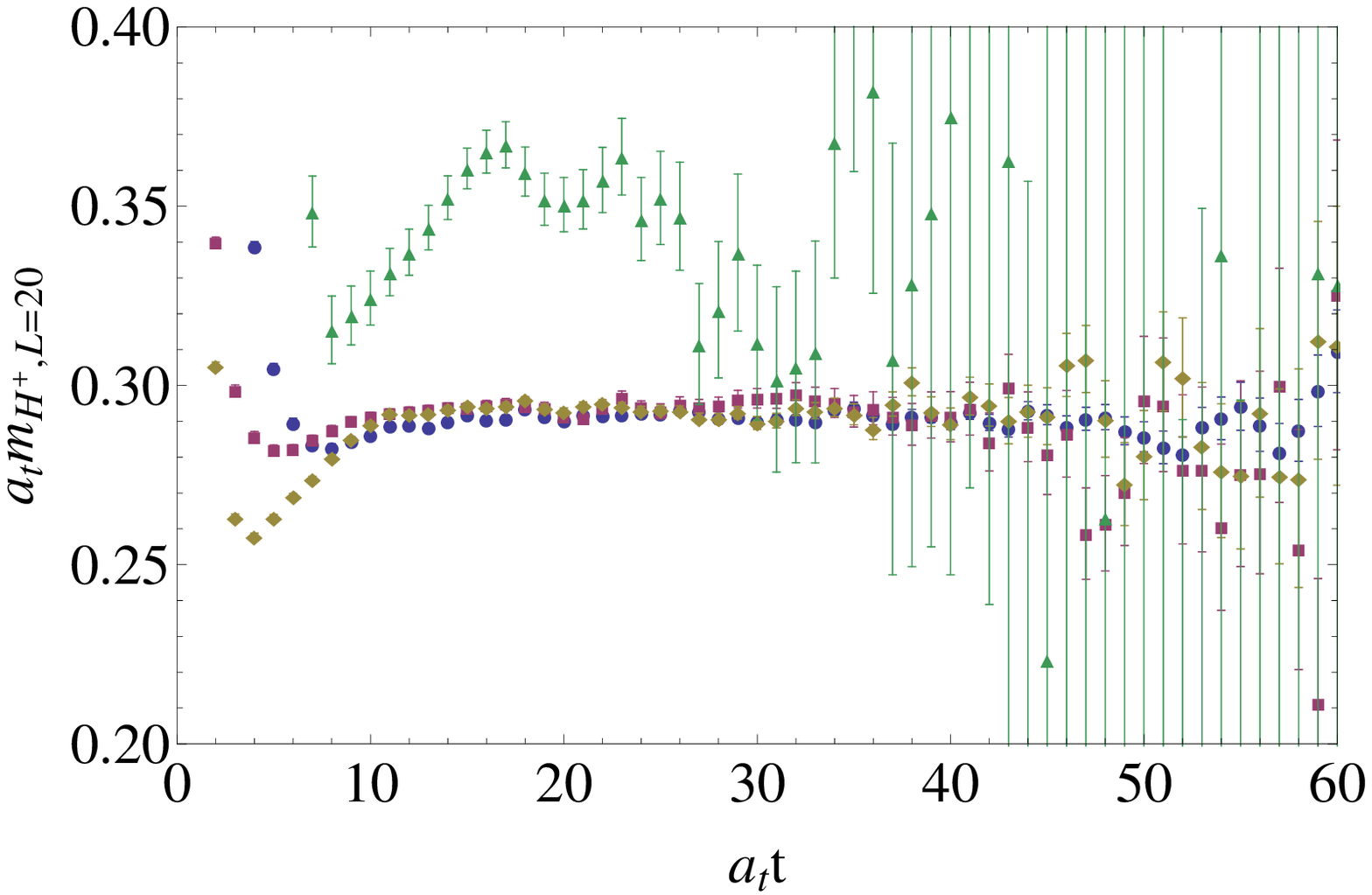}}}
\caption{(Color online) Effective mass plots for the four different $H^+$ embedding combinations calculated using the $20^3\times256$ lattices, using (a) the pure SS data and (b) a matrix-Prony rotation about timeslice $\tau=15$. The four embedding combinations shown are $H^{+}_{11}$ (blue circles), $H^{+}_{12}$ (red squares), $H^{+}_{21}$ (brown diamonds), and $H^{+}_{22}$ (green triangles).
\label{fig:H embeds}}
\end{figure}

As discussed in \sect{two_omega_lattice}, the strangeness -3 $H^{+}$ irrep has a dominant overlap with a spin-3/2 particle, whose ground state is the $\Omega^-$ particle. This irrep has two embeddings, making for a total of four possible source/sink embedding combinations. We will use the notation $H^{+}_{ij}$ for correlation functions with source embedding type $i$ and sink embedding type $j$. In \Fig{H embeds}, these four combinations are plotted for both the pure SS data and for data that has undergone a matrix-Prony rotation at timeslice $\tau=15$ for the $20^3\times256$ lattices.
While the largest overlap is observed in the $H^{+}_{11}$ combination, it is apparent from the figure that the highest embedding combination, $H^{+}_{22}$, has a significant amount of excited state contamination. Furthermore, this embedding combination does not have enough overlap with the ground state for a signal to appear before the onset of the baryonic noise around timeslice $t=40$. This lack of overlap with the ground state persists even following the matrix-Prony rotation, indicating that this higher embedding combination has an overlap with the ground state that cannot be resolved with the statistics available for this calculation. A similar situation is observed for the $32^3\times256$ lattices.

\begin{figure}[!ht]
\centering
\mbox{
\subfigure[]{\includegraphics[width=.5\columnwidth,angle=0]{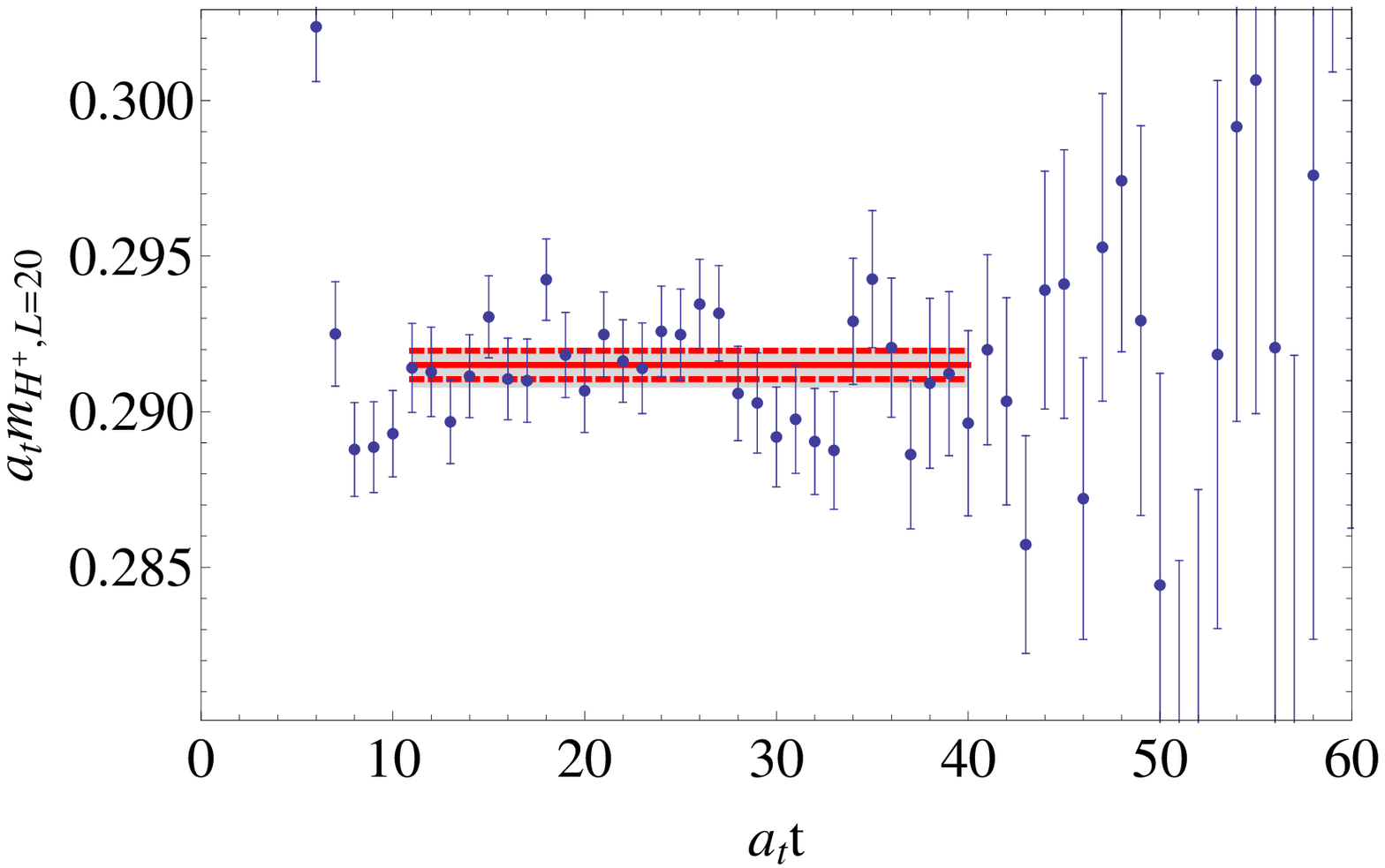}}\quad\subfigure[]{\includegraphics[width=.5\columnwidth,angle=0]{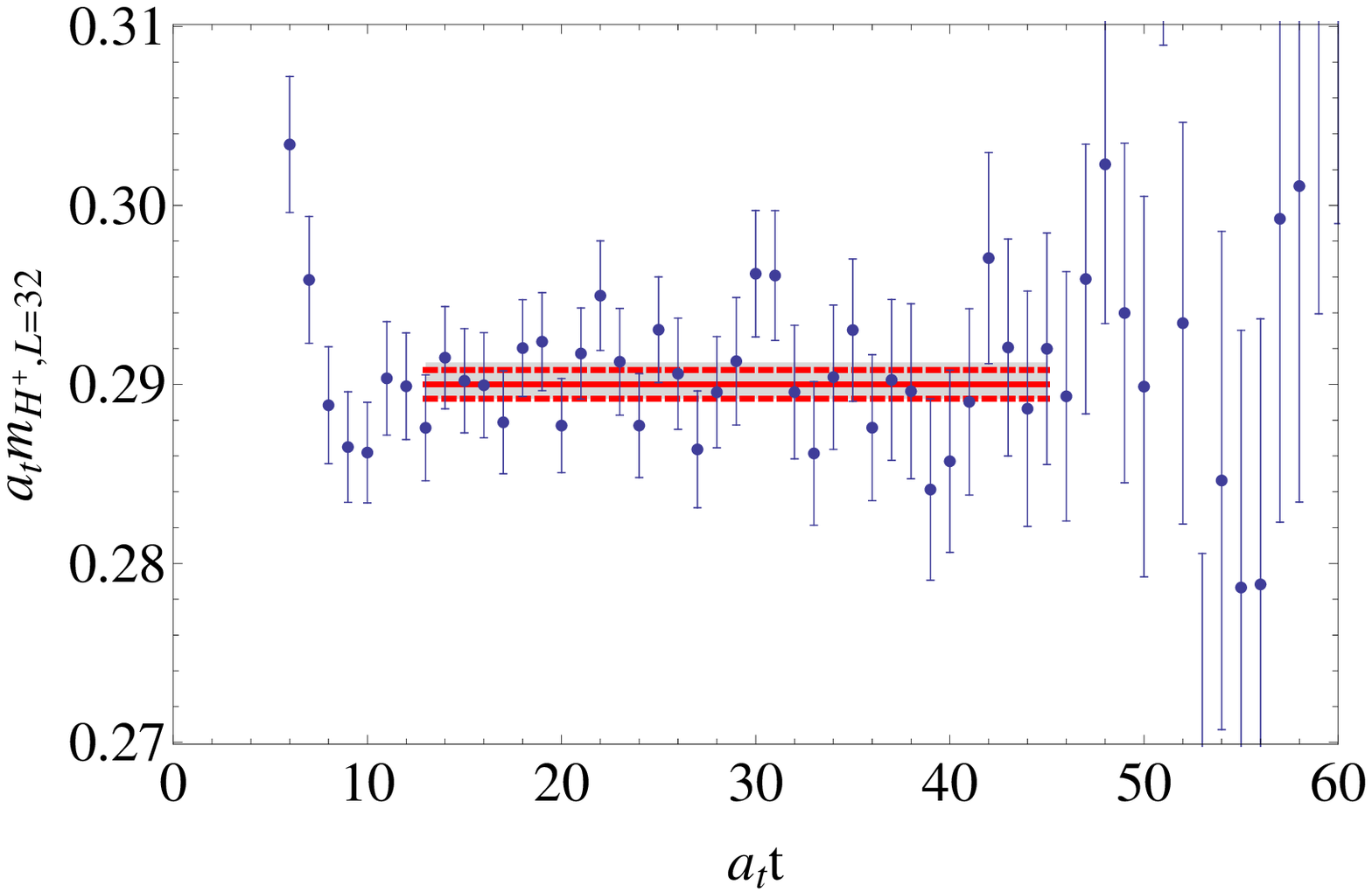}}}
\caption{(Color online) Effective mass plots for the $H^+$ (S=$\frac{3}{2}$) $\Omega^-$ baryon calculated using (a) $20^3\times256$ and (b) $32^3\times256$ lattices. The fit value is the solid red line, with statistical uncertainties the dashed red line. The grey box is the statistical plus the systematic uncertainties. The fit values are shown in \Table{H values}.
\label{fig:H masses}}
\end{figure}

Given the high number of measurements made for this calculation, this finding throws into doubt the utility of $H_{22}$ for any method attempting a better extraction of the $\Omega^-$ ground state. Specifically worrisome is the possibility that the contribution of even small amounts of this correlator to an effective mass plateau may lead to an inaccurately high ground state energy. As such, this embedding combination is removed from consideration in the following $H^{+}$ discussion. The remaining three embedding combinations have significant overlap with the ground state, and the effective mass plot for the sum of the bootstrap ensembles of the three lowest embedding combinations is shown in \Fig{H masses} for both the $20^3\times256$ and the $32^3\times256$ lattices. The results of the fits to each data set are shown in \Table{H values}.

\begin{table}[!ht]
\caption{\label{tab:H values}{Fit values for $H^{+}$ system energy levels (in dimensionless units, $a_{t}E$). }}
\begin{ruledtabular}
\begin{tabular}{c|cc|ccccc}
Irrep & Lattice Size & & $a_{t}E$ & $\sigma_{E,stat.}$ & $\sigma_{E,sys.}$ & $\chi^2$/dof & Q \\\hline
$H^{+}$ & $20^3\times256$ & & 0.291501 & 0.000457 & $^{+0.000099}_{-0.000268}$ & 1.003 & 0.460\\
 & $32^3\times256$ & & 0.290001 & 0.000804 & $^{+0.000418}_{-0.000001}$ & 0.850 & 0.708\\
\end{tabular}
\end{ruledtabular}
\end{table}

Using the values in \Table{H values} one can determine that the difference in energy between the two volumes is $\delta{E}_{H^{+}}=0.00150\pm0.00105$, where the statistical and systematic errors have been combined in quadrature. This sub-percent level difference is indicative of very small volume effects in the calculation. Also, if the most accurate data from the $20^3\times256$ lattices is used to naively set the scale, the resulting spatial lattice spacing would be $a_s=0.12038\pm0.00022$ fm, a percent-level difference from the scale set in \refcite{Lin:2008pr}, reflecting small differences from the physical point extrapolation.

\subsection{$s=\frac{1}{2}$: $G_{1}^{+}$ Irrep} \label{sec:G1+}         
\begin{figure}[!ht]
\centering
\mbox{
\includegraphics[width=.8\columnwidth,angle=0]{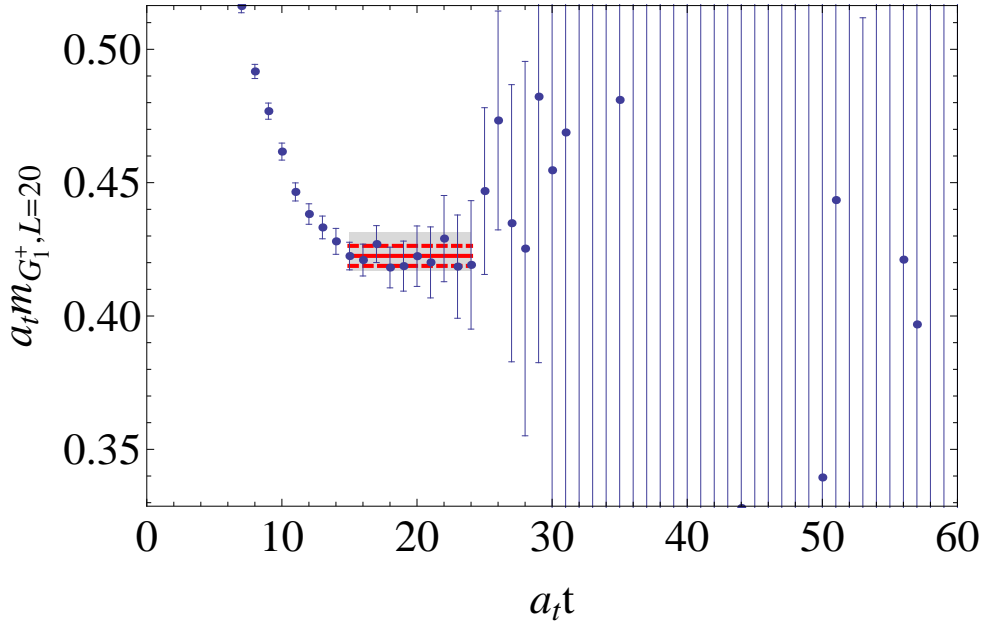}}
\caption{(Color online) Effective mass plot for the $G_1^+$ (S=$\frac{1}{2}$) $\Omega^-$ baryon calculated using $20^3\times256$ lattice. The fit value is the solid red line, with statistical uncertainties the dashed red line. The grey box is the statistical plus the systematic uncertainties. The fit value is tabulated in \Table{G values}.
\label{fig:G1 masses}}
\end{figure}

On the $20^3\times256$ lattice size calculations were also performed on the $G_{1}^{+}$ irrep, which has only one embedding. This is an excited state of the $\Omega^-$ particle with $S=\frac{1}{2}$, with the effective mass shown in \Fig{G1 masses} and the results of fitting the plateau in \Table{G values}. The results in \Fig{G1 masses} clearly show worse signal to noise behavior than for the $H^{+}$ state. This behavior precluded the examination of the two-$G_{1}^{+}$ system. The ratio of the extracted $G_{1}^{+}$ mass to the $H^{+}$ mass lattices compares within error to that extracted in \refcite{Bulava:2010yg}.
\begin{table}[!ht]
\caption{\label{tab:G values}{Fit values for $G_{1}^{+}$ system energy levels (in dimensionless units, $a_{t}E$).}}
\begin{ruledtabular}
\begin{tabular}{c|cc|ccccc}
Irrep & Lattice Size & & $a_{t}E$ & $\sigma_{E,stat.}$ & $\sigma_{E,sys.}$ & $\chi^2$/dof & Q \\\hline
$G_{1}^{+}$ & $20^3\times256$ & & 0.422541 & 0.003754 & $^{+0.005010}_{-0.002036}$ & 0.409 & 0.931\\
\end{tabular}
\end{ruledtabular}
\end{table}

\section{Strangeness -6 systems}\label{sec:two_omega_results}

\subsection{$s=\left(\frac{3}{2}\otimes\frac{3}{2}\right)$: The $A_1^{+}$, $E^{+}$, and $T_2^{+}$ Irreps}
Two strangeness -3 $H^{+}$ baryons (two $\Omega^-$ particles) can combine to make a strangeness -6 system. By forcing this system to be in a relative s-wave the angular momentum of the resulting state will be entirely determined by the spin combinations allowed, which for the two $\Omega^-$ system are the $S=0$ (the $A_1^{+}$ irrep) and the $S=2$ (the $E^{+}$ and $T_2^{+}$ irreps) angular momentum states.

\begin{figure}[!ht]
\centering
\mbox{
\subfigure[]{\includegraphics[width=.5\columnwidth,angle=0]{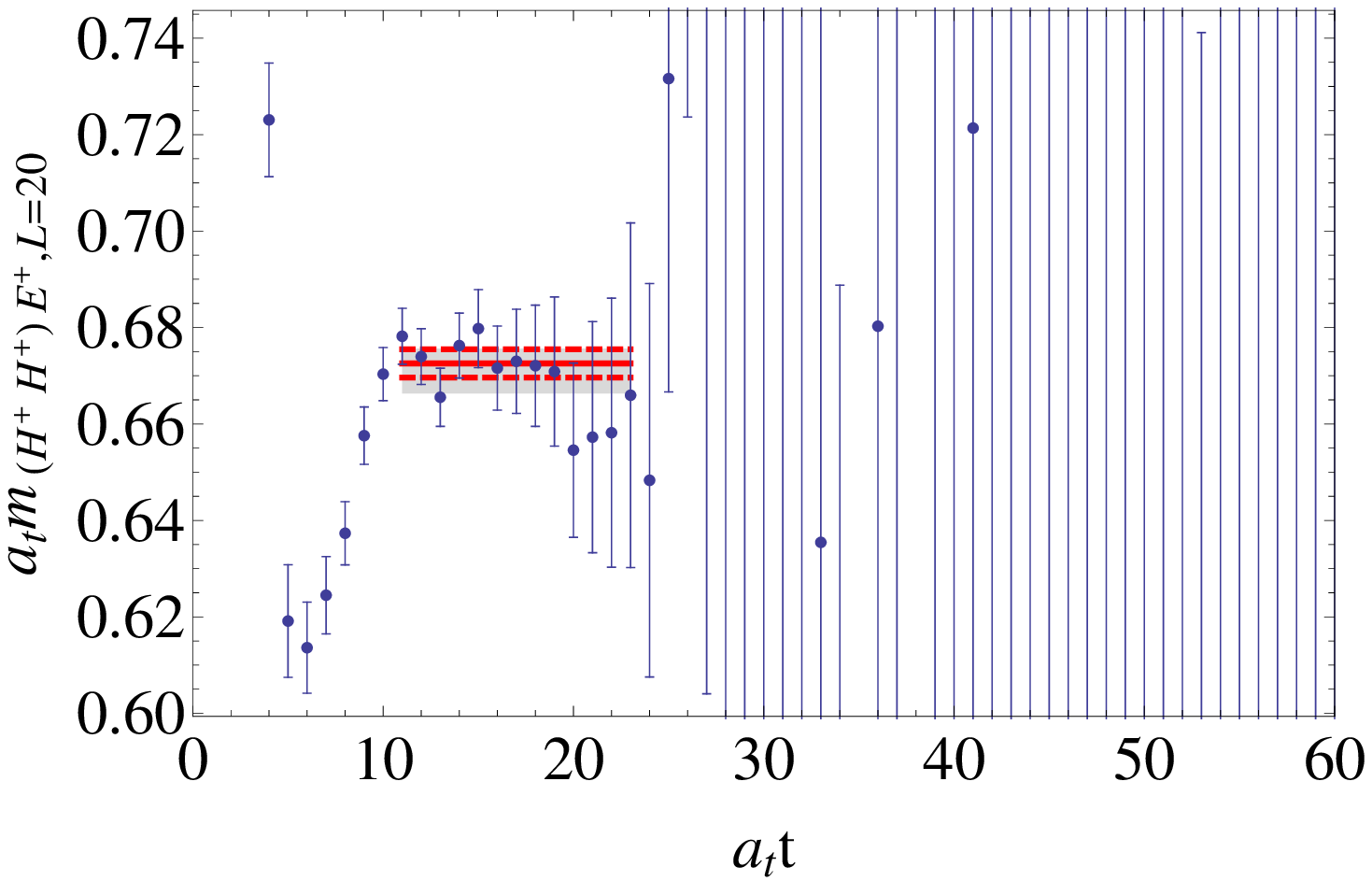}}\quad\subfigure[]{\includegraphics[width=.5\columnwidth,angle=0]{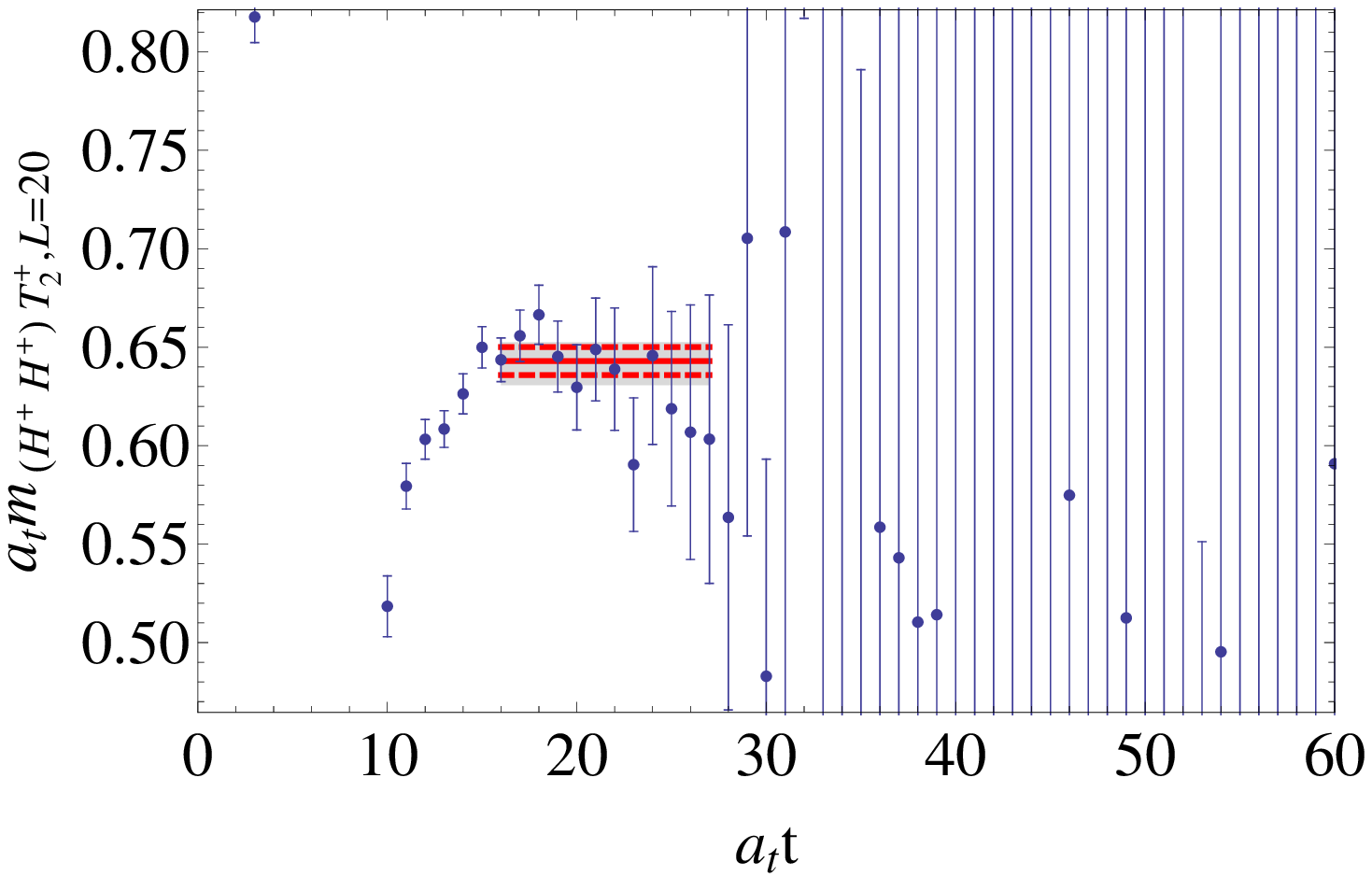}}}
\caption{(Color online) Effective mass plots of the (a) $E^{+}$ and (b) $T_2^{+}$ irreps (the $S=2$ two $\Omega^-$ system) calculated using the $20^3\times256$ lattice. The fit value is the solid red line, with statistical uncertainties the dashed red line. The grey box is the statistical plus the systematic uncertainties. The fit values are tabulated in \Table{S2 values}.
\label{fig:HH_ET2 masses}}
\end{figure}

For the $A_1^{+}$ irrep each of the $H^{+}$ baryons can be put into two embeddings at both the source and the sink, allowing for six embedding combinations: $A_{1;11,11}^{+}$, $A_{1;11,22}^{+}$, $A_{1;12,11}^{+}$, $A_{1;12,22}^{+}$, $A_{1;22,11}^{+}$, and $A_{1;22,22}^{+}$ where $\Gamma_{ij,kl}$ has source embeddings $i$ and $j$ with sink embeddings $k$ and $l$ for irrep $\Gamma$. Through a similar analysis of each embedding combination as was performed for the $H^{+}$, the combinations $A_{1;12,22}^{+}$ and $A_{1;22,22}^{+}$ were observed to plateau well above the common ground state that the other combinations found. Note that these two are the only combinations where it is impossible to avoid a contribution from contractions similar to those found in the $H^{+}_{22}$ embedding combination, and thus they likely suffer from a similar set of excited state contaminations. Given these observations, these two embedding combinations are excluded from the analysis of the $A_1^{+}$ system. The remaining embedding combinations are summed and result in the effective mass plots in \Fig{HH_ET2 masses} for both the $20^3\times256$ and the $32^3\times256$ lattices, with the fit values and energy shifts ($\Delta{E}$ as defined in \eq{E_shift}) given in \Table{S0 values}.
\begin{figure}[!ht]
\centering
\mbox{
\subfigure[]{\includegraphics[width=.5\columnwidth,angle=0]{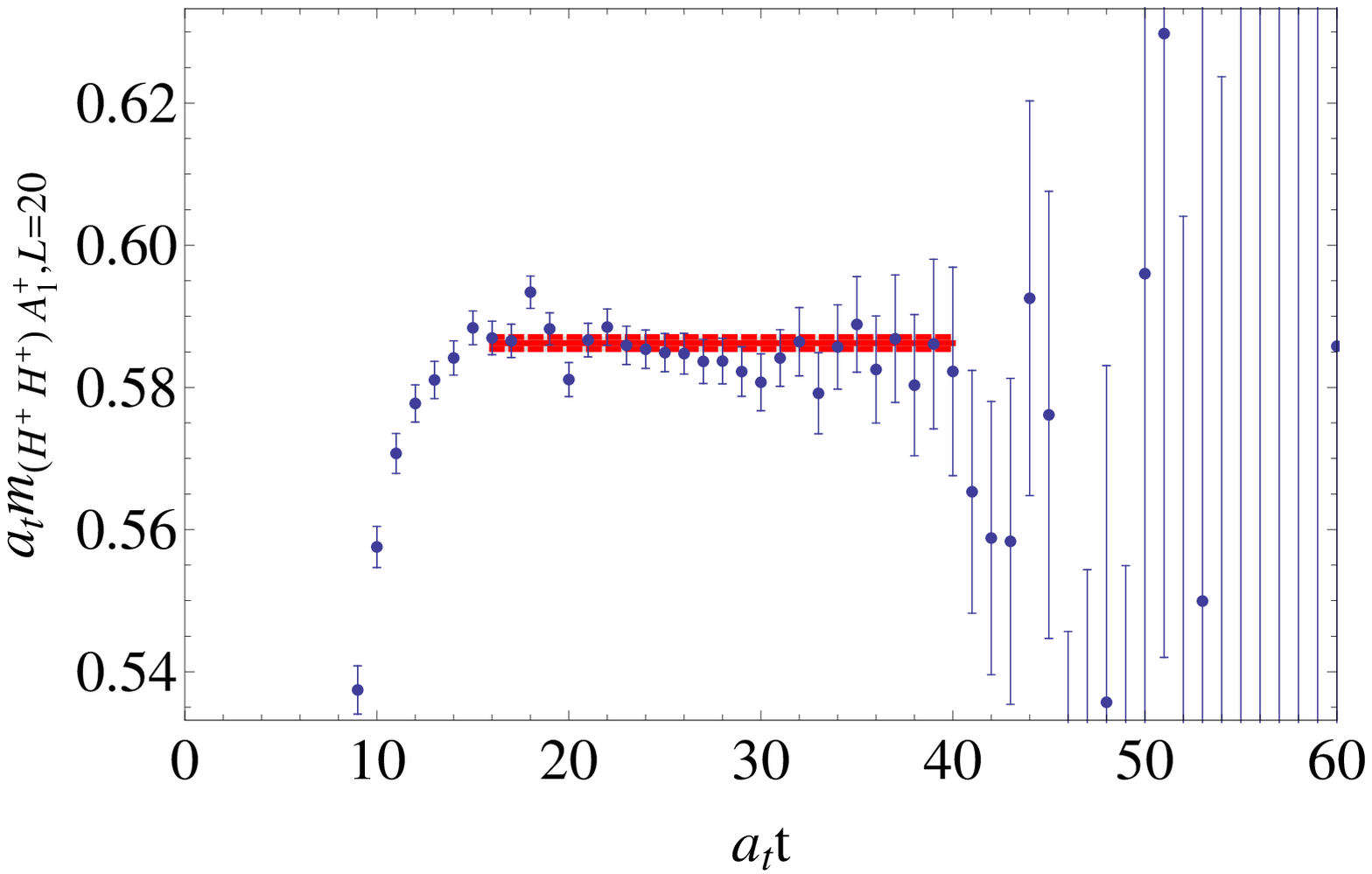}}\quad\subfigure[]{\includegraphics[width=.5\columnwidth,angle=0]{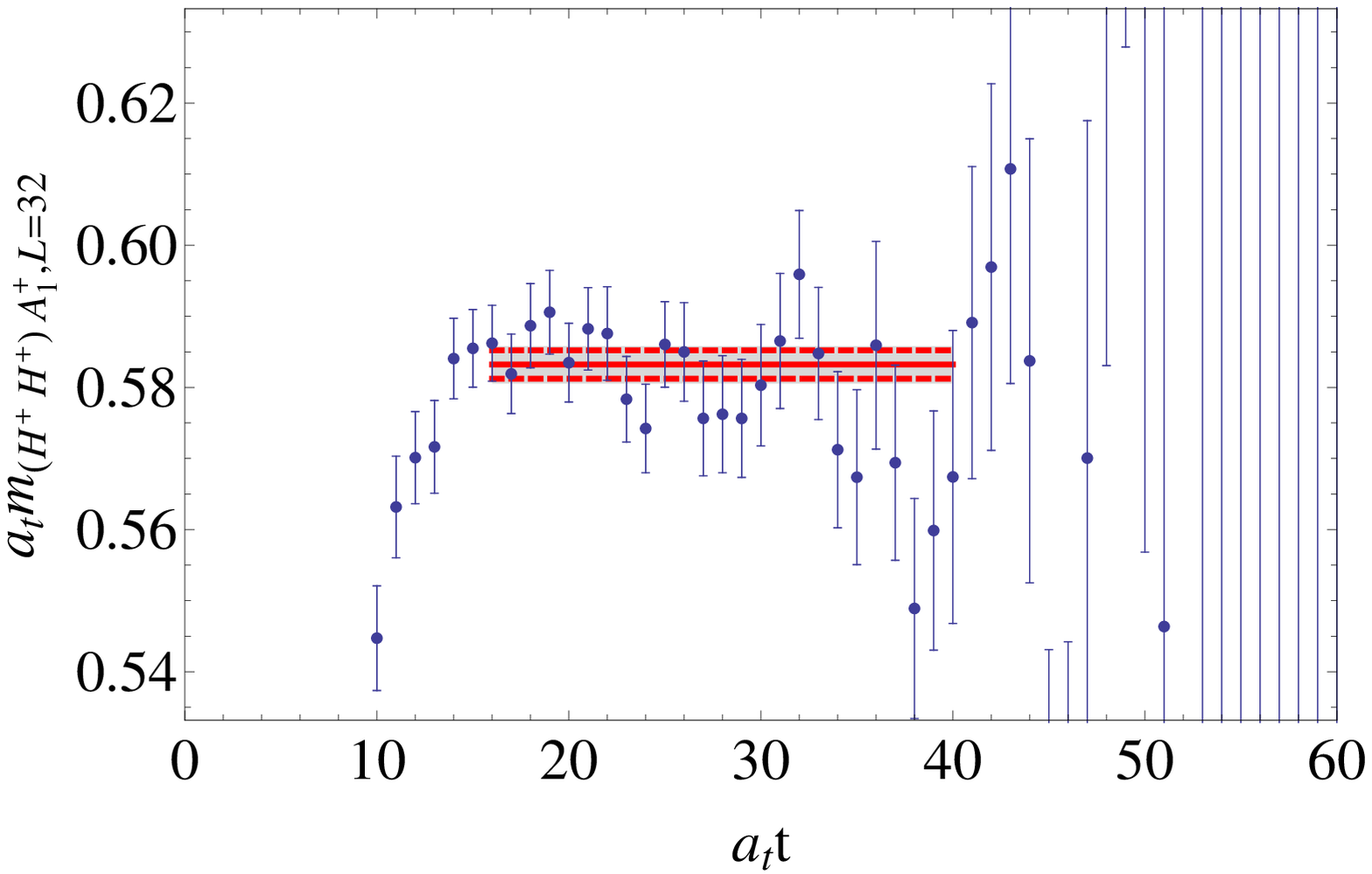}}}
\caption{(Color online) Effective mass plots for the $A_1^+$ (S=0) two $\Omega^-$ system calculated using (a) $20^3\times256$ and (b) $32^3\times256$ lattices. The fit value is the solid red line, with statistical uncertainties the dashed red line. The grey box is the statistical plus the systematic uncertainties. The fit values are shown in \Table{S0 values}.
\label{fig:HH_A1 masses}}
\end{figure}
\begin{table}[!ht]
\caption{\label{tab:S0 values}{Fit values and Energy Shifts for $A_1^{+}$ system energy levels (in dimensionless units, $a_{t}E$).}}
\begin{ruledtabular}
\begin{tabular}{c|c|ccccccc}
Irrep & Lattice Size & $a_{t}E$ & $\sigma_{E,stat.}$ & $\sigma_{E,sys.}$ & $\chi^2$/dof & Q & $a_{t}\Delta{E}$ & $\sigma_{\Delta{E},stat.}$\\\hline
$A_1^{+}$ & $20^3\times256$ & 0.586235 & 0.000843 & $^{+0.000091}_{-0.000348}$ & 1.105 & 0.327 & 0.00323 & 0.00124\\
 & $32^3\times256$ & 0.583224 & 0.002002 & $^{+0.000577}_{-0.000680}$ & 1.086 & 0.350 & 0.00322 & 0.00257\\
\end{tabular}
\end{ruledtabular}
\end{table}

The $E^{+}$ and $T_2^{+}$ irreps correspond to the $S=2$ two $\Omega^-$ system. Due to the limited number of quark spin states available to make this spin structure, the embedding combinations that will produce a non-zero result are much more limited than in the $A_1^{+}$ case. Specifically, placing either both of the source or both of the sink baryons into the first $H^{+}$ embedding is forbidden. Furthermore, the more complicated structure inherent to this spin state significantly reduced the signal to noise ratio in many of the remaining embedding combinations, leaving usable signals only the combinations $E^{+}_{12,22}$ and $T^{+}_{2;12,22}$.
The effective mass plots for these embeddings are shown in \Fig{HH_ET2 masses} and the fit values and energy shifts are in \Table{S2 values}. From \Table{S2 values} one can see that the two irreps achieve statistically separate lowest energy states, despite coupling to states with the same set of quantum numbers. This would indicate that at least one, and possibly both irreps, are failing to achieve the correct ground state of the $S=2$ two $\Omega^-$ system. In both cases, however, the states achieved are at a significantly higher energy level than for the $S=0$ case, implying a much more repulsive channel, as expected from Pauli exclusion arguments.
\begin{table}[!ht]
\caption{\label{tab:S2 values}{Fit values and Energy Shifts for the $S=2$ two $\Omega^-$ system energy levels (in dimensionless units, $a_{t}E$).}}
\begin{ruledtabular}
\begin{tabular}{c|c|ccccccc}
Irrep & Lattice Size & $a_{t}E$ & $\sigma_{E,stat.}$ & $\sigma_{E,sys.}$ & $\chi^2$/dof & Q & $a_{t}\Delta{E}$ & $\sigma_{\Delta{E},stat.}$\\\hline
$T_2^{+}$ & $20^3\times256$ & 0.642961 & 0.007136 & $^{+0.002502}_{-0.005120}$ & 0.925 & 0.514 & 0.05996 & 0.00719\\
$E^{+}$ & $20^3\times256$ & 0.67256 & 0.00293 & $^{+0.00013}_{-0.00329}$ & 0.500 & 0.916 & 0.08956 & 0.00307\\
\end{tabular}
\end{ruledtabular}
\end{table}

\subsection{$S=\left(\frac{3}{2}\otimes\frac{1}{2}\right)$: The $T_1^{+}$ and $E^{+}$ Irreps}
\begin{figure}[!ht]
\centering
\mbox{
\subfigure[]{\includegraphics[width=.5\columnwidth,angle=0]{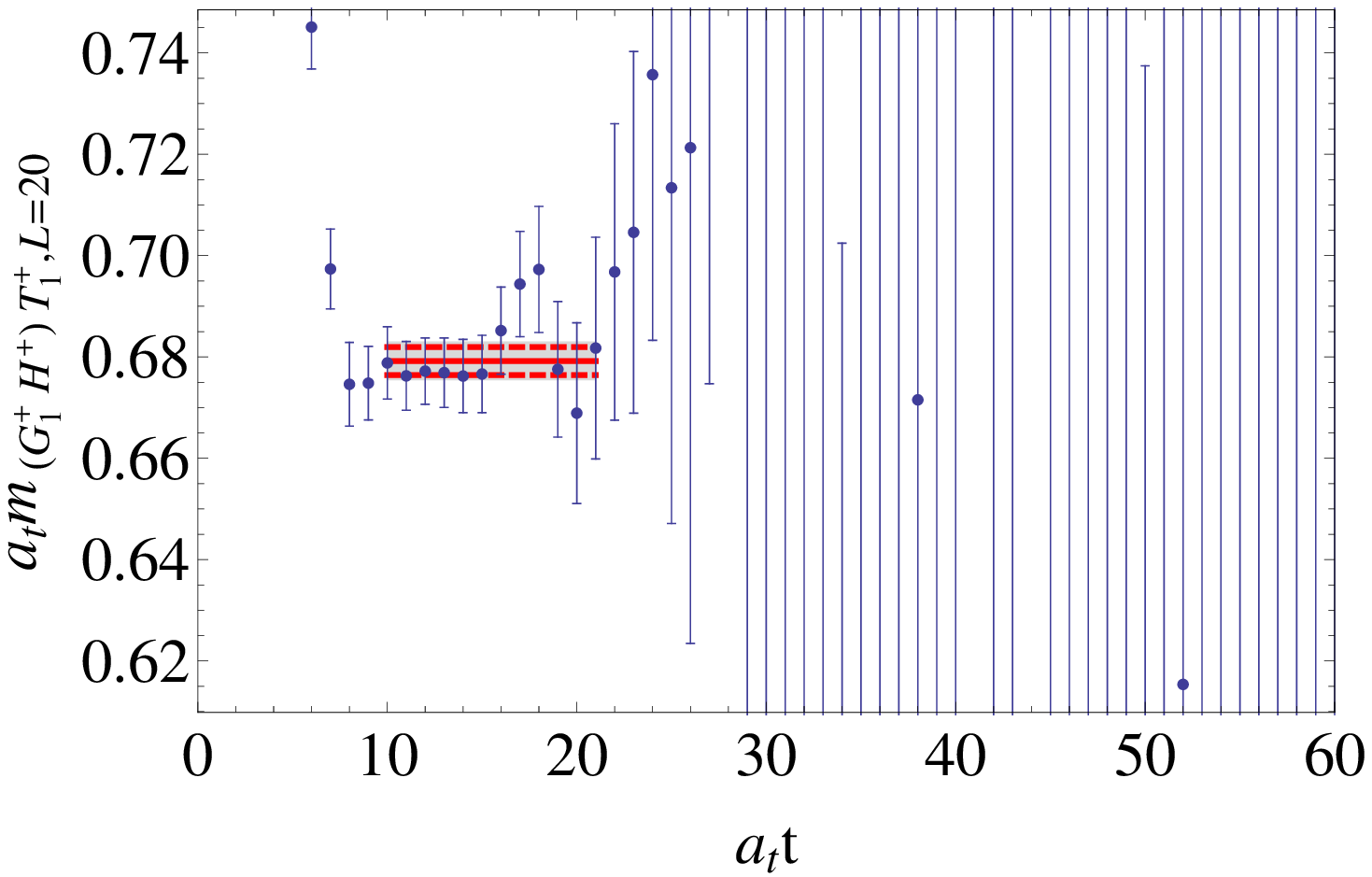}}\quad\subfigure[]{\includegraphics[width=.5\columnwidth,angle=0]{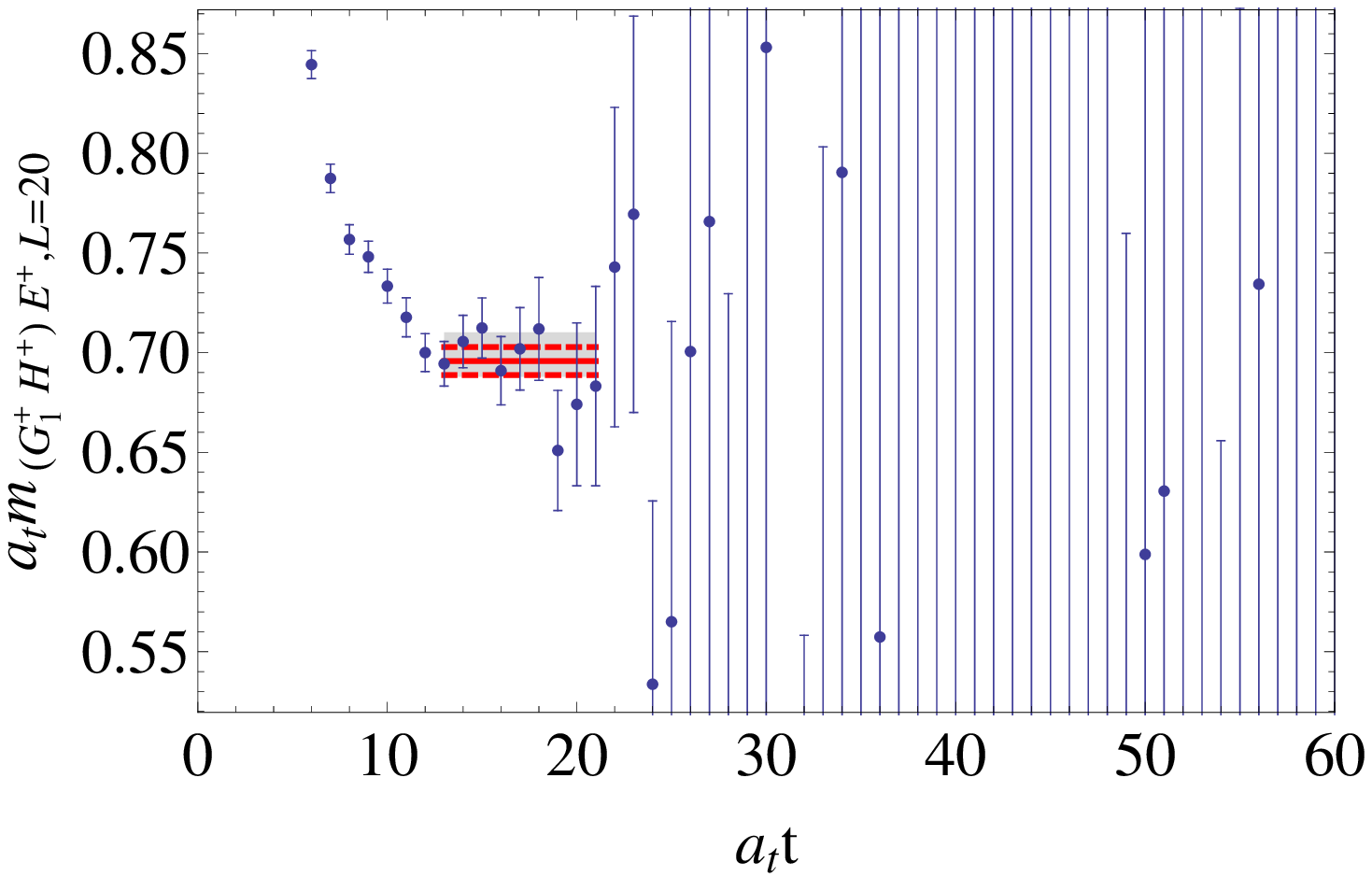}}}
\caption{(Color online) Effective mass plots of the $\left(\frac{3}{2}\otimes\frac{1}{2}\right)$-coupled  two $\Omega^-$ system calculated using the $20^3\times256$ lattice for (a) $T_1^+$ (S=1) and (b) $E^+$ (S=2) irreps. The fit value is the solid red line, with statistical uncertainties the dashed red line. The grey box is the statistical plus the systematic uncertainties. The fit values are tabulated in \Table{GH values}.
\label{fig:GH_ET1 masses}}
\end{figure}

The final system examined in the calculation was the interaction between two strangeness -3 baryons, one in the $H^{+}$ irrep and one in the $G_1^{+}$ irrep, on the $20^3\times256$ lattices. This system is unique in that it involves explicitly placing one baryon in in the $S=\frac{1}{2}$ excited state. The final spin combinations allowed for this combination are $S=1$, which will fall into the $T_1^{+}$ irrep, and $S=2$ which will again fall into either the $E^{+}$ or $T_2^{+}$ irreps. The $G_1^{+}$ irrep has only one embedding, so the possible embedding combinations are determined solely by the remaining $H^{+}$ irrep baryon. In the $S=1$, $T_1^{+}$ case usable signals were recovered for embedding combinations with the source $H^{+}$ in the first embedding and the sink in the second. For the $S=2$ case a signal was uncovered only for the $E^{+}$ irrep again with the source $H^{+}$ in the first embedding and the sink in the second. The effective mass plots for each are shown in \Fig{GH_ET1 masses} and the fit results and energy shifts are in \Table{GH values}.
\begin{table}[!ht]
\caption{\label{tab:GH values}{Fit values and Energy Shifts for the $G_1^{+}\otimes{H}^{+}$ system energy levels (in dimensionless units, $a_{t}E$).}}
\begin{ruledtabular}
\begin{tabular}{c|c|ccccccc}
Irrep & Lattice Size & $a_{t}E$ & $\sigma_{E,stat.}$ & $\sigma_{E,sys.}$ & $\chi^2$/dof & Q & $a_{t}\Delta{E}$ & $\sigma_{\Delta{E},stat.}$\\\hline
$T_1^{+}$ & $20^3\times256$ & 0.679179 & 0.002773 & $^{+0.001087}_{-0.000992}$ & 0.389 & 0.961 & -0.03486 & 0.00469\\
$E^{+}$ & $20^3\times256$ & 0.695768 & 0.007049 & $^{+0.007353}_{-0.000774}$ & 0.747 & 0.650 & -0.01827 & 0.00800\\
\end{tabular}
\end{ruledtabular}
\end{table}

\section{Scattering and $k\cdot\text{cot}\delta$}\label{sec:kcotdelta}
With energy levels of each baryon and system of baryons determined in \sect{single_omega_results} and \sect{two_omega_results}, respectively, one can now extract scattering information from this data following the discussion from \sect{two_omega_lattice}. Returning first to the $\left(\frac{3}{2}\otimes\frac{3}{2}\right)$ $S=0$ two $\Omega^-$ system, the data from two different volumes will allow for two applications of \eq{Phase_Shift} and, in combination with \eq{effrange}, an extraction of the scattering length $a$. In principle the range parameter $r$ will also be extracted, however that term in \eq{effrange} will also be contaminated by contributions from all of the higher order terms in the expansion, and as such will be unreliable. This determination of the scattering characteristics of the two $\Omega$ system will allow for definitive statements to be made on the form of interaction between these baryons in light of the conflicting claims of \refcite{Zhang:2000sv} and \refcite{Wang:1995bg}.

\begin{figure}[!ht]
\centering
\mbox{
\subfigure[]{\includegraphics[width=.5\columnwidth,angle=0]{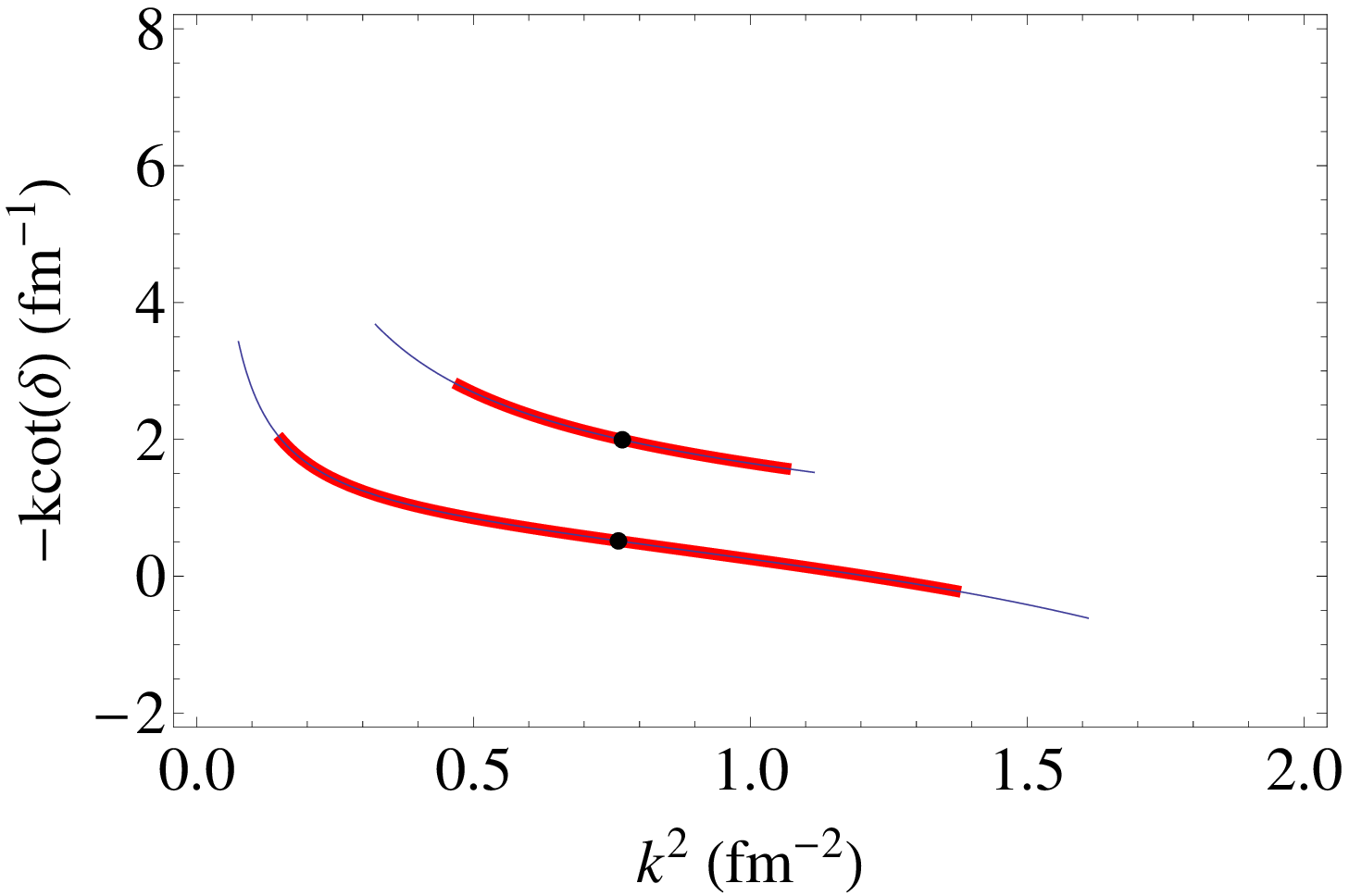}}\quad\subfigure[]{\includegraphics[width=.46\columnwidth,angle=0]{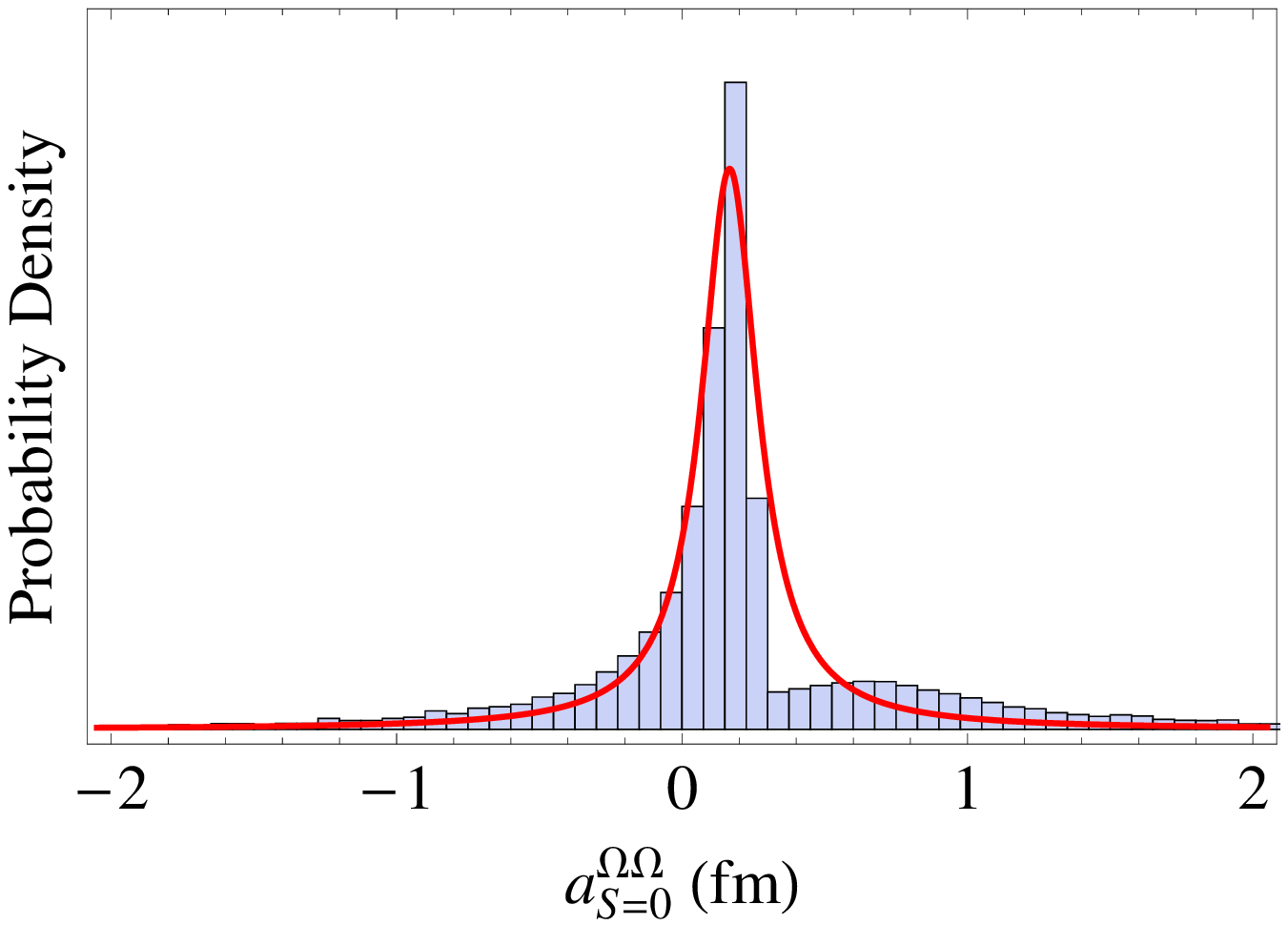}}}
\caption{(Color online) Plot of (a) $k\text{cot}\delta$ and (b) the distribution of scattering lengths $a$ for the $S=0$ two $\Omega^-$ system. The $k\text{cot}\delta$ plot consists of the central value (black dot), statistical error (thick inner red line), and statistical plus systematic error (thin outer blue line) for the $20^3\times256$ lattices (top line) and the $32^3\times256$ lattices (bottom line). The scattering length distribution has the probability density histogram overlaid with a Lorentz distribution fit to the data. This distribution has a central peak at 0.16 fm with a 68\% confidence interval (equivalent to 1$\sigma$) of 0.22 fm.
\label{fig:S0kcotdelta}}
\end{figure}

Using the data from \Table{H values} and \Table{S0 values} along with \eq{E_shift} and \eq{Phase_Shift} one can determine the $k^2$ and $k\cot \delta(k)$ values for the $S=0$ two $\Omega^-$ system. The (dimensionful) results are shown in \Fig{S0kcotdelta}(a) along with the systematic and statistical errors. To obtain the scattering length with a correct propagation of errors, a distribution of the parameter $a$ is generated. To accomplish this a series of 10,000 pairs of random values were taken from the distributions of $k^2$ for both the $20^3\times256$ and the $32^3\times256$ lattice data. These distributions were normal distributions defined from the mean value and statistical plus systematic uncertainty of $k^2$ obtained from \Table{H values}, \Table{S0 values}, and \eq{E_shift}. Each random pair was then used to generate a pair of $k\text{cot}\delta$ values using \eq{Phase_Shift}, which was then fit to \eq{effrange} to produce a single value for the scattering length $a$. The distribution of these values is shown in \Fig{S0kcotdelta}(b), with a resulting two $\Omega^-$ scattering length in the S=0 channel of
\beq
a^{\Omega\Omega}_{S=0}=0.16 \pm 0.22 \ \text{fm}.
\eeq
Note that the distribution fit to the data is that of a Lorentz distribution and not a normal distribution, due to the specific form that \eq{S} takes. If one used standard error propagation techniques to determine the uncertainty in the scattering length, the form of the distribution as a Lorentz distribution rather than a normal distribution would result in the quotation of too large an uncertainty.

From \Fig{S0kcotdelta}(a) one can see that the central value of $k^2$ observed does not change appreciably between the two different volumes. Also, in \Fig{S0kcotdelta}(b) the distribution of the extracted scattering length is strongly peaked at very small values. Both of these pieces of information are indicative of a very weakly repulsive system. Indeed, if one operates with the assumption of natural sizes for the range and higher order parameters in the effective range expansion then the Lorentz distribution for the scattering length would provide an 79.5\% chance that the system is repulsive and a 20.5\% chance that it is attractive. Additionally, the $\Delta E$ values in \Table{S0 values} are positive (repulsive) and small within the $1 \sigma$ error band for both $20^3\times256$ and $32^3\times256$ lattices.   Thus, from our current lattice calculations, we find evidence that the system is consistent with the weakly repulsive scenario in \refcite{Wang:1995bg} and inconsistent with the deeply bound state found in \refcite{Zhang:2000sv}.  Ultimately, more calculations are required to acquire a full error budget of the systematic, but these systematics for the $\Omega\Omega$ system are not expected to be appreciable for the reasons mentioned earlier.

\begin{figure}[!ht]
\centering
\mbox{
\includegraphics[width=.9\columnwidth,angle=0]{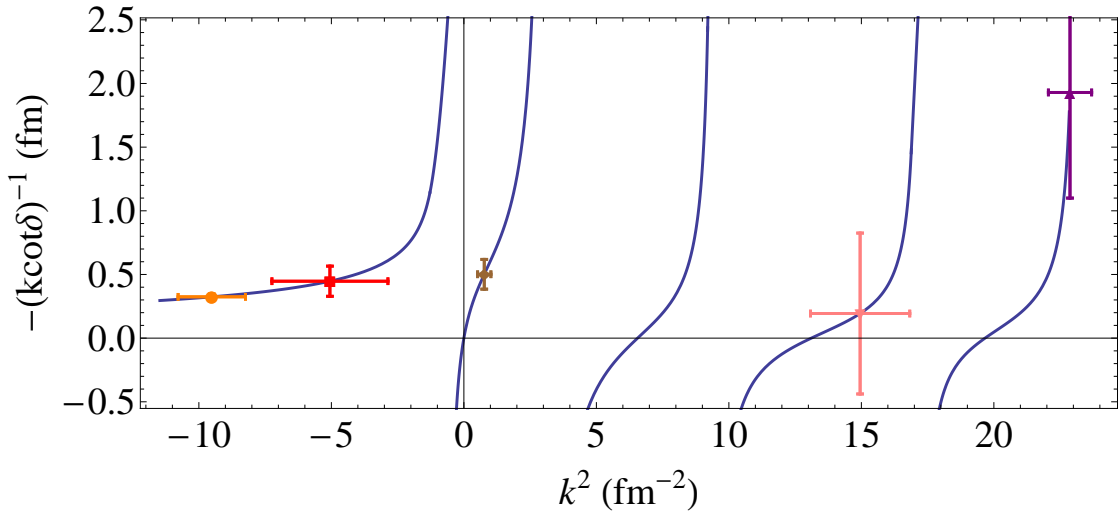}}
\caption{(Color online) Combined results for $k\text{cot}\delta$ for all two baryon systems studied on the $20^3\times256$ lattices. These systems are (from left to right) the $S=1$ $G_1^{+}H^{+}$ system (orange circle), the $S=2$ $G_1^{+}H^{+}$ system (red square), the $S=0$ $H^{+}H^{+}$ system (brown diamond), the $S=2$ $H^{+}H^{+}$ system in the $T_2^{+}$ irrep (pink downward triangle), and the $S=2$ $H^{+}H^{+}$ system in the $E^{+}$ irrep (purple upward triangle). The errors shown are statistical only.
\label{fig:kcotdelta}}
\end{figure}

Finally, for the other two baryon systems studied on the $20^3\times256$ lattices a calculation of $k^2$ and $k\text{cot}\delta$ can be made in a manner similar to that for the $S=0$ two $\Omega^-$ system detailed above. However, as these calculations were performed on only one lattice size, a reliable scattering length cannot be obtained and one can only quote the inverse of $k\text{cot}\delta$ as a proxy for the scattering length. The inverse of $k\text{cot}\delta$ is plotted for each of the two baryon systems studied on the $20^3\times256$ lattices in \Fig{kcotdelta}. The $S=0$ two $\Omega^-$ system already discussed is the middle of the five points shown.

Of note in \Fig{kcotdelta} is that the $G_1^{+}H^{+}$ excited state system appears to be attractive in both the $S=1$ and the $S=2$ channels. However, without additional lattice volumes one cannot make a claim as to whether these states would be bound or simply scatter attractively (resonances). Also, in nature the $G_1^{+}$ state would decay strongly to a $\Xi$ baryon and a kaon in a relative p-wave with the release of several hundred MeV of energy (assuming a $G_1^{+}$ mass of 2500 MeV\cite{Bulava:2010yg}), a far lower energy arrangement than any putative $G_1^{+}H^{+}$ bound state. Also, the ordering of the $G_1^{+}H^{+}$ systems does seem to follow the general trend observed and expected in the other two $\Omega^-$ system calculated where the lower spin state results in a more attractive/less repulsive interaction.

For the $S=2$ $H^{+}H^{+}$ system in \Fig{kcotdelta}, the two different irrep calculations clearly obtain two different states. It is notable that the two states lie on different branches of the $k\text{cot}\delta$ function, and that neither are on the principle branch that the $S=0$ states are on. This would indicate both that the $S=2$ states are strongly repulsive (which one could qualitatively predict from the Pauli exclusion principle) and that it is possible that neither of the extracted states are actually the $S=2$ ground state. If the interaction is repulsive enough then the volume in which the objects are contained may not be able to physically contain the two baryon system, leading the scattering interaction to be pushed to higher branches of $k\text{cot}\delta$. In a regime where the interaction fits within the lattice volume, the interaction energy should scale inversely with the volume, with any stronger scaling indicating a volume that is too small. A calculation of the $S=2$ $H^{+}H^{+}$ system at a second volume would point toward a resolution of this question.

\section{Conclusion} \label{sec:concl}         
We have performed the initial s-wave lattice calculations of the two spin-3/2 $\Omega^-$ system for $S$=0 and 2 channels, as well as spin-$\frac{1}{2}$/spin-$\frac{3}{2}$ coupled two-$\Omega$ system for $S$=1 and 2 channels. Using lattice configurations with pion mass $m_\pi\sim390 \ \text{MeV}$,  calculations were performed in a volume with $L$=2.5 fm, and in the $S=0$ case, also with $L$=3.9 fm.  The $S$=0 results demonstrate that the two-$\Omega$ system in this channel is most likely weakly repulsive, with a scattering length
\beq
a^{\Omega\Omega}_{S=0}=0.16 \pm 0.22 \ \text{fm}.
\eeq
Also, the energy of interaction was positive and small within errors for both $L$=2.5 fm and $L$=3.9 fm, whose values are states in \Table{S0 values}.  As our calculations rely on no phenomenology, we assert that these findings should provide significant evidence supporting the conclusion of a weakly repulsive system in \refcite{Wang:1995bg}, as opposed to a deeply bound state in \refcite{Zhang:2000sv}. These results also provide an interesting complement to previous studies\cite{Beane:2009py,Beane:2010hg,Inoue:2010es,Beane:2011iw} of hyperon interactions, where many of the interactions have been found to be attractive and contain bound states at a pion mass of 390 MeV. The difference between the evidence for other bound hyperon states and the conclusion in this work of a weakly repulsive $\Omega\Omega$ state may simply reflect a much stronger influence of light-quark dynamics in the valence sector of the $\Lambda\Lambda$ and $\Xi\Xi$ systems. Further studies at different pion masses approaching the physical point are needed to gain a better understanding of the similarities and differences of each of these systems.

Central to this work was the use of interpolating operators of definite lattice cubic symmetries, which allowed us to look at exotic channels of the two-$\Omega$ system.  In some of these channels we find evidence for an attractive interaction.  Future studies with larger statistics and multiple volumes will better elucidate the behavior of two-$\Omega$ systems in these channels.  The minimal light-quark dependence of the $\Omega$ system implies that the results presented in this work should be `near-physical'. Extension of this work to configurations with lighter pion mass, as well as the development of an $\chi$EFT for this system, will quantify this statement.

\begin{acknowledgments}
We thank  P. Bedaque, M. Cheng, W. Haxton, B. Joo, B. Tiburzi, P. Vranas, M. J. Savage,  S. Wallace, and A. Walker-Loud for many useful discussions. Using Chroma\cite{Edwards:2004sx}, the configurations used were generated on uBGL while the propagator inversions and contractions were performed on the Edge cluster with the QUDA GPU library\cite{Clark:2009wm}, both at LLNL. We are indebted to B. Joo for his help in implementing the QUDA libraries on Edge.  This work was performed under the auspices of the U.S. Department of Energy by LLNL under Contract No. DE-AC52-07NA27344 and the UNEDF SciDAC Grant No. DE-FC02-07ER41457.  This research was partially supported by the LLNL LDRD ``Unlocking the Universe with High Performance Computing" 10-ERD-033 and by the LLNL Multiprogrammatic and Institutional Computing program through a Tier 1 Grand Challenge award.
\end{acknowledgments}

\bibliography{omega_omega} %

\begin{thebibliography}{41}
\expandafter\ifx\csname natexlab\endcsname\relax\def\natexlab#1{#1}\fi
\expandafter\ifx\csname bibnamefont\endcsname\relax
  \def\bibnamefont#1{#1}\fi
\expandafter\ifx\csname bibfnamefont\endcsname\relax
  \def\bibfnamefont#1{#1}\fi
\expandafter\ifx\csname citenamefont\endcsname\relax
  \def\citenamefont#1{#1}\fi
\expandafter\ifx\csname url\endcsname\relax
  \def\url#1{\texttt{#1}}\fi
\expandafter\ifx\csname urlprefix\endcsname\relax\def\urlprefix{URL }\fi
\providecommand{\bibinfo}[2]{#2}
\providecommand{\eprint}[2][]{\url{#2}}

\bibitem[{\citenamefont{Luscher}(1986)}]{Luscher:1986pf}
\bibinfo{author}{\bibfnamefont{M.}~\bibnamefont{Luscher}},
  \bibinfo{journal}{Commun. Math. Phys.} \textbf{\bibinfo{volume}{105}},
  \bibinfo{pages}{153} (\bibinfo{year}{1986}).

\bibitem[{\citenamefont{Luscher}(1991)}]{Luscher:1990ux}
\bibinfo{author}{\bibfnamefont{M.}~\bibnamefont{Luscher}},
  \bibinfo{journal}{Nucl.Phys.} \textbf{\bibinfo{volume}{B354}},
  \bibinfo{pages}{531} (\bibinfo{year}{1991}).

\bibitem[{\citenamefont{Yamazaki et~al.}(2004)}]{Yamazaki:2004qb}
\bibinfo{author}{\bibfnamefont{T.}~\bibnamefont{Yamazaki}} \bibnamefont{et~al.}
  (\bibinfo{collaboration}{CP-PACS}), \bibinfo{journal}{Phys. Rev.}
  \textbf{\bibinfo{volume}{D70}}, \bibinfo{pages}{074513}
  (\bibinfo{year}{2004}), \eprint{hep-lat/0402025}.

\bibitem[{\citenamefont{Beane et~al.}(2006)\citenamefont{Beane, Bedaque,
  Orginos, and Savage}}]{Beane:2005rj}
\bibinfo{author}{\bibfnamefont{S.~R.} \bibnamefont{Beane}},
  \bibinfo{author}{\bibfnamefont{P.~F.} \bibnamefont{Bedaque}},
  \bibinfo{author}{\bibfnamefont{K.}~\bibnamefont{Orginos}}, \bibnamefont{and}
  \bibinfo{author}{\bibfnamefont{M.~J.} \bibnamefont{Savage}}
  (\bibinfo{collaboration}{NPLQCD}), \bibinfo{journal}{Phys. Rev.}
  \textbf{\bibinfo{volume}{D73}}, \bibinfo{pages}{054503}
  (\bibinfo{year}{2006}), \eprint{hep-lat/0506013}.

\bibitem[{\citenamefont{Beane et~al.}(2008)}]{Beane:2007xs}
\bibinfo{author}{\bibfnamefont{S.~R.} \bibnamefont{Beane}}
  \bibnamefont{et~al.}, \bibinfo{journal}{Phys. Rev.}
  \textbf{\bibinfo{volume}{D77}}, \bibinfo{pages}{014505}
  (\bibinfo{year}{2008}), \eprint{0706.3026}.

\bibitem[{\citenamefont{Feng et~al.}(2010)\citenamefont{Feng, Jansen, and
  Renner}}]{Feng:2009ij}
\bibinfo{author}{\bibfnamefont{X.}~\bibnamefont{Feng}},
  \bibinfo{author}{\bibfnamefont{K.}~\bibnamefont{Jansen}}, \bibnamefont{and}
  \bibinfo{author}{\bibfnamefont{D.~B.} \bibnamefont{Renner}},
  \bibinfo{journal}{Phys. Lett.} \textbf{\bibinfo{volume}{B684}},
  \bibinfo{pages}{268} (\bibinfo{year}{2010}), \eprint{0909.3255}.

\bibitem[{\citenamefont{Dudek et~al.}(2011{\natexlab{a}})\citenamefont{Dudek,
  Edwards, Peardon, Richards, and Thomas}}]{Dudek:2010ew}
\bibinfo{author}{\bibfnamefont{J.~J.} \bibnamefont{Dudek}},
  \bibinfo{author}{\bibfnamefont{R.~G.} \bibnamefont{Edwards}},
  \bibinfo{author}{\bibfnamefont{M.~J.} \bibnamefont{Peardon}},
  \bibinfo{author}{\bibfnamefont{D.~G.} \bibnamefont{Richards}},
  \bibnamefont{and} \bibinfo{author}{\bibfnamefont{C.~E.}
  \bibnamefont{Thomas}}, \bibinfo{journal}{Phys.Rev.}
  \textbf{\bibinfo{volume}{D83}}, \bibinfo{pages}{071504}
  (\bibinfo{year}{2011}{\natexlab{a}}), \eprint{1011.6352}.

\bibitem[{\citenamefont{Beane et~al.}(2011{\natexlab{a}})}]{Beane:2011sc}
\bibinfo{author}{\bibfnamefont{S.}~\bibnamefont{Beane}} \bibnamefont{et~al.}
  (\bibinfo{collaboration}{NPLQCD Collaboration})
  (\bibinfo{year}{2011}{\natexlab{a}}), \eprint{1107.5023}.

\bibitem[{\citenamefont{Yagi et~al.}(2011)\citenamefont{Yagi, Hashimoto,
  Morimatsu, and Ohtani}}]{Yagi:2011jn}
\bibinfo{author}{\bibfnamefont{T.}~\bibnamefont{Yagi}},
  \bibinfo{author}{\bibfnamefont{S.}~\bibnamefont{Hashimoto}},
  \bibinfo{author}{\bibfnamefont{O.}~\bibnamefont{Morimatsu}},
  \bibnamefont{and} \bibinfo{author}{\bibfnamefont{M.}~\bibnamefont{Ohtani}}
  (\bibinfo{year}{2011}), \eprint{1108.2970}.

\bibitem[{\citenamefont{Beane et~al.}(2010{\natexlab{a}})}]{Beane:2010hg}
\bibinfo{author}{\bibfnamefont{S.~R.} \bibnamefont{Beane}} \bibnamefont{et~al.}
  (\bibinfo{collaboration}{NPLQCD}) (\bibinfo{year}{2010}{\natexlab{a}}),
  \eprint{1012.3812}.

\bibitem[{\citenamefont{Inoue et~al.}(2011{\natexlab{a}})}]{Inoue:2010es}
\bibinfo{author}{\bibfnamefont{T.}~\bibnamefont{Inoue}} \bibnamefont{et~al.}
  (\bibinfo{collaboration}{HAL QCD}), \bibinfo{journal}{Phys. Rev. Lett.}
  \textbf{\bibinfo{volume}{106}}, \bibinfo{pages}{162002}
  (\bibinfo{year}{2011}{\natexlab{a}}), \eprint{1012.5928}.

\bibitem[{\citenamefont{Beane et~al.}(2011{\natexlab{b}})}]{Beane:2011iw}
\bibinfo{author}{\bibfnamefont{S.~R.} \bibnamefont{Beane}} \bibnamefont{et~al.}
  (\bibinfo{collaboration}{NPLQCD}) (\bibinfo{year}{2011}{\natexlab{b}}),
  \eprint{1109.2889}.

\bibitem[{\citenamefont{Dudek et~al.}(2010)\citenamefont{Dudek, Edwards,
  Peardon, Richards, and Thomas}}]{Dudek:2010wm}
\bibinfo{author}{\bibfnamefont{J.~J.} \bibnamefont{Dudek}},
  \bibinfo{author}{\bibfnamefont{R.~G.} \bibnamefont{Edwards}},
  \bibinfo{author}{\bibfnamefont{M.~J.} \bibnamefont{Peardon}},
  \bibinfo{author}{\bibfnamefont{D.~G.} \bibnamefont{Richards}},
  \bibnamefont{and} \bibinfo{author}{\bibfnamefont{C.~E.}
  \bibnamefont{Thomas}}, \bibinfo{journal}{Phys. Rev.}
  \textbf{\bibinfo{volume}{D82}}, \bibinfo{pages}{034508}
  (\bibinfo{year}{2010}), \eprint{1004.4930}.

\bibitem[{\citenamefont{Dudek et~al.}(2011{\natexlab{b}})}]{Dudek:2011tt}
\bibinfo{author}{\bibfnamefont{J.~J.} \bibnamefont{Dudek}} \bibnamefont{et~al.}
  (\bibinfo{year}{2011}{\natexlab{b}}), \eprint{1102.4299}.

\bibitem[{\citenamefont{Basak et~al.}(2007)}]{Basak:2007kj}
\bibinfo{author}{\bibfnamefont{S.}~\bibnamefont{Basak}} \bibnamefont{et~al.},
  \bibinfo{journal}{Phys. Rev.} \textbf{\bibinfo{volume}{D76}},
  \bibinfo{pages}{074504} (\bibinfo{year}{2007}), \eprint{0709.0008}.

\bibitem[{\citenamefont{Bulava et~al.}(2010)}]{Bulava:2010yg}
\bibinfo{author}{\bibfnamefont{J.}~\bibnamefont{Bulava}} \bibnamefont{et~al.},
  \bibinfo{journal}{Phys. Rev.} \textbf{\bibinfo{volume}{D82}},
  \bibinfo{pages}{014507} (\bibinfo{year}{2010}), \eprint{1004.5072}.

\bibitem[{\citenamefont{Edwards et~al.}(2011)\citenamefont{Edwards, Dudek,
  Richards, and Wallace}}]{Edwards:2011jj}
\bibinfo{author}{\bibfnamefont{R.~G.} \bibnamefont{Edwards}},
  \bibinfo{author}{\bibfnamefont{J.~J.} \bibnamefont{Dudek}},
  \bibinfo{author}{\bibfnamefont{D.~G.} \bibnamefont{Richards}},
  \bibnamefont{and} \bibinfo{author}{\bibfnamefont{S.~J.}
  \bibnamefont{Wallace}} (\bibinfo{year}{2011}), \eprint{1104.5152}.

\bibitem[{\citenamefont{Peardon et~al.}(2009)}]{Peardon:2009gh}
\bibinfo{author}{\bibfnamefont{M.}~\bibnamefont{Peardon}} \bibnamefont{et~al.}
  (\bibinfo{collaboration}{Hadron Spectrum}), \bibinfo{journal}{Phys. Rev.}
  \textbf{\bibinfo{volume}{D80}}, \bibinfo{pages}{054506}
  (\bibinfo{year}{2009}), \eprint{0905.2160}.

\bibitem[{\citenamefont{Clark et~al.}(2010)\citenamefont{Clark, Babich, Barros,
  Brower, and Rebbi}}]{Clark:2009wm}
\bibinfo{author}{\bibfnamefont{M.~A.} \bibnamefont{Clark}},
  \bibinfo{author}{\bibfnamefont{R.}~\bibnamefont{Babich}},
  \bibinfo{author}{\bibfnamefont{K.}~\bibnamefont{Barros}},
  \bibinfo{author}{\bibfnamefont{R.~C.} \bibnamefont{Brower}},
  \bibnamefont{and} \bibinfo{author}{\bibfnamefont{C.}~\bibnamefont{Rebbi}},
  \bibinfo{journal}{Comput. Phys. Commun.} \textbf{\bibinfo{volume}{181}},
  \bibinfo{pages}{1517} (\bibinfo{year}{2010}), \eprint{0911.3191}.

\bibitem[{\citenamefont{Basak et~al.}(2005)}]{Basak:2005ir}
\bibinfo{author}{\bibfnamefont{S.}~\bibnamefont{Basak}} \bibnamefont{et~al.}
  (\bibinfo{collaboration}{Lattice Hadron Physics (LHPC)}),
  \bibinfo{journal}{Phys. Rev.} \textbf{\bibinfo{volume}{D72}},
  \bibinfo{pages}{074501} (\bibinfo{year}{2005}), \eprint{hep-lat/0508018}.

\bibitem[{\citenamefont{Jaffe}(1977)}]{Jaffe:1976yi}
\bibinfo{author}{\bibfnamefont{R.~L.} \bibnamefont{Jaffe}},
  \bibinfo{journal}{Phys. Rev. Lett.} \textbf{\bibinfo{volume}{38}},
  \bibinfo{pages}{195} (\bibinfo{year}{1977}).

\bibitem[{\citenamefont{Golak et~al.}(1997)}]{Golak:1996hj}
\bibinfo{author}{\bibfnamefont{J.}~\bibnamefont{Golak}} \bibnamefont{et~al.},
  \bibinfo{journal}{Phys. Rev.} \textbf{\bibinfo{volume}{C55}},
  \bibinfo{pages}{2196} (\bibinfo{year}{1997}), \eprint{nucl-th/9612065}.

\bibitem[{\citenamefont{Nogga et~al.}(2002)\citenamefont{Nogga, Kamada, and
  Gloeckle}}]{Nogga:2001ef}
\bibinfo{author}{\bibfnamefont{A.}~\bibnamefont{Nogga}},
  \bibinfo{author}{\bibfnamefont{H.}~\bibnamefont{Kamada}}, \bibnamefont{and}
  \bibinfo{author}{\bibfnamefont{W.}~\bibnamefont{Gloeckle}},
  \bibinfo{journal}{Phys. Rev. Lett.} \textbf{\bibinfo{volume}{88}},
  \bibinfo{pages}{172501} (\bibinfo{year}{2002}), \eprint{nucl-th/0112060}.

\bibitem[{\citenamefont{Kamimura et~al.}(2009)\citenamefont{Kamimura, Kino, and
  Hiyama}}]{Kamimura:2008fx}
\bibinfo{author}{\bibfnamefont{M.}~\bibnamefont{Kamimura}},
  \bibinfo{author}{\bibfnamefont{Y.}~\bibnamefont{Kino}}, \bibnamefont{and}
  \bibinfo{author}{\bibfnamefont{E.}~\bibnamefont{Hiyama}},
  \bibinfo{journal}{Prog.Theor.Phys.} \textbf{\bibinfo{volume}{121}},
  \bibinfo{pages}{1059} (\bibinfo{year}{2009}), \eprint{0809.4772}.

\bibitem[{\citenamefont{Zhang et~al.}(2000)\citenamefont{Zhang, Yu, Ching, Ho,
  and Lu}}]{Zhang:2000sv}
\bibinfo{author}{\bibfnamefont{Z.~Y.} \bibnamefont{Zhang}},
  \bibinfo{author}{\bibfnamefont{Y.~W.} \bibnamefont{Yu}},
  \bibinfo{author}{\bibfnamefont{C.~R.} \bibnamefont{Ching}},
  \bibinfo{author}{\bibfnamefont{T.~H.} \bibnamefont{Ho}}, \bibnamefont{and}
  \bibinfo{author}{\bibfnamefont{Z.-D.} \bibnamefont{Lu}},
  \bibinfo{journal}{Phys. Rev.} \textbf{\bibinfo{volume}{C61}},
  \bibinfo{pages}{065204} (\bibinfo{year}{2000}).

\bibitem[{\citenamefont{Wang et~al.}(1995)\citenamefont{Wang, Ping, Wu, Teng,
  and Goldman}}]{Wang:1995bg}
\bibinfo{author}{\bibfnamefont{F.}~\bibnamefont{Wang}},
  \bibinfo{author}{\bibfnamefont{J.-l.} \bibnamefont{Ping}},
  \bibinfo{author}{\bibfnamefont{G.-h.} \bibnamefont{Wu}},
  \bibinfo{author}{\bibfnamefont{L.-j.} \bibnamefont{Teng}}, \bibnamefont{and}
  \bibinfo{author}{\bibfnamefont{J.~T.} \bibnamefont{Goldman}},
  \bibinfo{journal}{Phys. Rev.} \textbf{\bibinfo{volume}{C51}},
  \bibinfo{pages}{3411} (\bibinfo{year}{1995}), \eprint{nucl-th/9512014}.

\bibitem[{\citenamefont{Page and Reddy}(2006)}]{Page:2006ud}
\bibinfo{author}{\bibfnamefont{D.}~\bibnamefont{Page}} \bibnamefont{and}
  \bibinfo{author}{\bibfnamefont{S.}~\bibnamefont{Reddy}},
  \bibinfo{journal}{Ann. Rev. Nucl. Part. Sci.} \textbf{\bibinfo{volume}{56}},
  \bibinfo{pages}{327} (\bibinfo{year}{2006}), \eprint{astro-ph/0608360}.

\bibitem[{\citenamefont{Lin et~al.}(2009)}]{Lin:2008pr}
\bibinfo{author}{\bibfnamefont{H.-W.} \bibnamefont{Lin}} \bibnamefont{et~al.}
  (\bibinfo{collaboration}{Hadron Spectrum}), \bibinfo{journal}{Phys. Rev.}
  \textbf{\bibinfo{volume}{D79}}, \bibinfo{pages}{034502}
  (\bibinfo{year}{2009}), \eprint{0810.3588}.

\bibitem[{\citenamefont{Beane et~al.}(2004)\citenamefont{Beane, Bedaque,
  Parreno, and Savage}}]{Beane:2003da}
\bibinfo{author}{\bibfnamefont{S.}~\bibnamefont{Beane}},
  \bibinfo{author}{\bibfnamefont{P.}~\bibnamefont{Bedaque}},
  \bibinfo{author}{\bibfnamefont{A.}~\bibnamefont{Parreno}}, \bibnamefont{and}
  \bibinfo{author}{\bibfnamefont{M.}~\bibnamefont{Savage}},
  \bibinfo{journal}{Phys.Lett.} \textbf{\bibinfo{volume}{B585}},
  \bibinfo{pages}{106} (\bibinfo{year}{2004}), \eprint{hep-lat/0312004}.

\bibitem[{\citenamefont{Beane et~al.}(2010{\natexlab{b}})}]{Beane:2009py}
\bibinfo{author}{\bibfnamefont{S.~R.} \bibnamefont{Beane}} \bibnamefont{et~al.}
  (\bibinfo{collaboration}{NPLQCD}), \bibinfo{journal}{Phys. Rev.}
  \textbf{\bibinfo{volume}{D81}}, \bibinfo{pages}{054505}
  (\bibinfo{year}{2010}{\natexlab{b}}), \eprint{0912.4243}.

\bibitem[{\citenamefont{Inoue et~al.}(2011{\natexlab{b}})}]{Inoue:2011ai}
\bibinfo{author}{\bibfnamefont{T.}~\bibnamefont{Inoue}} \bibnamefont{et~al.}
  (\bibinfo{collaboration}{HAL QCD Collaboration})
  (\bibinfo{year}{2011}{\natexlab{b}}), \eprint{1112.5926}.

\bibitem[{\citenamefont{Beane et~al.}(2009{\natexlab{a}})\citenamefont{Beane,
  Detmold, Luu, Orginos, Parreno et~al.}}]{Beane:2009gs}
\bibinfo{author}{\bibfnamefont{S.~R.} \bibnamefont{Beane}},
  \bibinfo{author}{\bibfnamefont{W.}~\bibnamefont{Detmold}},
  \bibinfo{author}{\bibfnamefont{T.~C.} \bibnamefont{Luu}},
  \bibinfo{author}{\bibfnamefont{K.}~\bibnamefont{Orginos}},
  \bibinfo{author}{\bibfnamefont{A.}~\bibnamefont{Parreno}},
  \bibnamefont{et~al.}, \bibinfo{journal}{Phys.Rev.}
  \textbf{\bibinfo{volume}{D80}}, \bibinfo{pages}{074501}
  (\bibinfo{year}{2009}{\natexlab{a}}), \eprint{0905.0466}.

\bibitem[{\citenamefont{Yamazaki et~al.}(2010)\citenamefont{Yamazaki,
  Kuramashi, Ukawa, and Collaboration}}]{Yamazaki:2009ua}
\bibinfo{author}{\bibfnamefont{T.}~\bibnamefont{Yamazaki}},
  \bibinfo{author}{\bibfnamefont{Y.}~\bibnamefont{Kuramashi}},
  \bibinfo{author}{\bibfnamefont{A.}~\bibnamefont{Ukawa}}, \bibnamefont{and}
  \bibinfo{author}{\bibfnamefont{f.~t. P.-C.} \bibnamefont{Collaboration}},
  \bibinfo{journal}{Phys.Rev.} \textbf{\bibinfo{volume}{D81}},
  \bibinfo{pages}{111504} (\bibinfo{year}{2010}), \eprint{0912.1383}.

\bibitem[{\citenamefont{Yuan et~al.}(1999)\citenamefont{Yuan, Zhang, Yu, and
  Shen}}]{Yuan:1999pg}
\bibinfo{author}{\bibfnamefont{X.~Q.} \bibnamefont{Yuan}},
  \bibinfo{author}{\bibfnamefont{Z.~Y.} \bibnamefont{Zhang}},
  \bibinfo{author}{\bibfnamefont{Y.~W.} \bibnamefont{Yu}}, \bibnamefont{and}
  \bibinfo{author}{\bibfnamefont{P.~N.} \bibnamefont{Shen}},
  \bibinfo{journal}{Phys. Rev.} \textbf{\bibinfo{volume}{C60}},
  \bibinfo{pages}{045203} (\bibinfo{year}{1999}), \eprint{nucl-th/9901069}.

\bibitem[{\citenamefont{Tiburzi and Walker-Loud}(2008)}]{Tiburzi:2008bk}
\bibinfo{author}{\bibfnamefont{B.~C.} \bibnamefont{Tiburzi}} \bibnamefont{and}
  \bibinfo{author}{\bibfnamefont{A.}~\bibnamefont{Walker-Loud}},
  \bibinfo{journal}{Phys. Lett.} \textbf{\bibinfo{volume}{B669}},
  \bibinfo{pages}{246} (\bibinfo{year}{2008}), \eprint{0808.0482}.

\bibitem[{\citenamefont{Ding and Liao}(2012)}]{Ding:2012sm}
\bibinfo{author}{\bibfnamefont{R.}~\bibnamefont{Ding}} \bibnamefont{and}
  \bibinfo{author}{\bibfnamefont{Y.}~\bibnamefont{Liao}}
  (\bibinfo{year}{2012}), \eprint{1201.0506}.

\bibitem[{\citenamefont{Maiani and Testa}(1990)}]{Maiani:1990ca}
\bibinfo{author}{\bibfnamefont{L.}~\bibnamefont{Maiani}} \bibnamefont{and}
  \bibinfo{author}{\bibfnamefont{M.}~\bibnamefont{Testa}},
  \bibinfo{journal}{Phys. Lett.} \textbf{\bibinfo{volume}{B245}},
  \bibinfo{pages}{585} (\bibinfo{year}{1990}).

\bibitem[{\citenamefont{Beane et~al.}(2009{\natexlab{b}})}]{Beane:2009kya}
\bibinfo{author}{\bibfnamefont{S.~R.} \bibnamefont{Beane}}
  \bibnamefont{et~al.}, \bibinfo{journal}{Phys. Rev.}
  \textbf{\bibinfo{volume}{D79}}, \bibinfo{pages}{114502}
  (\bibinfo{year}{2009}{\natexlab{b}}), \eprint{0903.2990}.

\bibitem[{\citenamefont{Luscher and Wolff}(1990)}]{Luscher:1990ck}
\bibinfo{author}{\bibfnamefont{M.}~\bibnamefont{Luscher}} \bibnamefont{and}
  \bibinfo{author}{\bibfnamefont{U.}~\bibnamefont{Wolff}},
  \bibinfo{journal}{Nucl.Phys.} \textbf{\bibinfo{volume}{B339}},
  \bibinfo{pages}{222} (\bibinfo{year}{1990}).

\bibitem[{\citenamefont{Bulava et~al.}(2009)\citenamefont{Bulava, Edwards,
  Engelson, Foley, Joo et~al.}}]{Bulava:2009jb}
\bibinfo{author}{\bibfnamefont{J.~M.} \bibnamefont{Bulava}},
  \bibinfo{author}{\bibfnamefont{R.~G.} \bibnamefont{Edwards}},
  \bibinfo{author}{\bibfnamefont{E.}~\bibnamefont{Engelson}},
  \bibinfo{author}{\bibfnamefont{J.}~\bibnamefont{Foley}},
  \bibinfo{author}{\bibfnamefont{B.}~\bibnamefont{Joo}}, \bibnamefont{et~al.},
  \bibinfo{journal}{Phys.Rev.} \textbf{\bibinfo{volume}{D79}},
  \bibinfo{pages}{034505} (\bibinfo{year}{2009}), \eprint{0901.0027}.

\bibitem[{\citenamefont{Edwards and Joo}(2005)}]{Edwards:2004sx}
\bibinfo{author}{\bibfnamefont{R.~G.} \bibnamefont{Edwards}} \bibnamefont{and}
  \bibinfo{author}{\bibfnamefont{B.}~\bibnamefont{Joo}}
  (\bibinfo{collaboration}{SciDAC}), \bibinfo{journal}{Nucl. Phys. Proc.
  Suppl.} \textbf{\bibinfo{volume}{140}}, \bibinfo{pages}{832}
  (\bibinfo{year}{2005}), \eprint{hep-lat/0409003}.

\end{thebibliography}

\end{document}